# CVD-graphene/graphene flakes dual-films as advanced DSSC counter electrodes

Andrea Capasso[1,†], Sebastiano Bellani[1,†], Alessandro Lorenzo Palma[2], Leyla Najafi[1,3], Antonio Esaù Del Rio Castillo[1], Nicola Curreli[1,4], Lucio Cinà[2], Vaidotas Miseikis[5], Camilla Coletti[5], Giuseppe Calogero[6], Vittorio Pellegrini[1,7], Aldo Di Carlo[2,8] and Francesco Bonaccorso[1,7,*]

[1] Graphene Labs, Istituto Italiano di Tecnologia, via Morego 30, 16163 Genova, Italy.
[2] C.H.O.S.E. (Centre for Hybrid and Organic Solar Energy), Department of Electronic Engineering, University of Rome Tor Vergata, Via del Politecnico 1, Rome 00133, Italy.
[3] Dipartimento di Chimica e Chimica Industriale, Università degli Studi di Genova, Via Dodecaneso 31, Genova, 16146, Italy.
[4] Department of Electrical and Electronic Engineering, University of Cagliari, P.zza d'Armi, 09123 Cagliari, Italy
[5] Center for Nanotechnology Innovation @ NEST, Istituto Italiano di Tecnologia, P.zza San Silvestro 12, Pisa, Italy.
[6] CNR-IPCF Istituto per i Processi Chimico-Fisici, via F. Stagno d'Alcontres 37, 98158 –Messina, Italy.
[7] BeDimensional Spa., Via Albisola 121, 16163 Genova, Italy.
[8] L.A.S.E. - Laboratory for Advanced Solar Energy, National University of Science and Technology "MISiS", 119049 Leninskiy prosect 6. Moscow, Russia.

[†] These authors equally contributed
*E-mail: francesco.bonaccorso@iit.it

## Abstract

The use of graphene-based electrodes is burgeoning in a wide range of applications, including solar cells, light emitting diodes, touch screens, field-effect transistors, photodetectors, sensors and energy storage systems. The success of such electrodes strongly depends on the implementation of effective production and processing methods for graphene. In this work, we take advantage of two different graphene production methods to design an advanced, conductive oxide- and platinum-free, graphene-based counter electrode for dye-sensitized solar cells (DSSCs). In particular, we exploit the combination of a graphene film, produced by chemical vapor deposition (CVD) (CVD-graphene), with few-layer graphene (FLG) flakes, produced by liquid phase exfoliation. The CVD-graphene is used as charge collector, while the FLG flakes, deposited atop by spray coating, act as catalyst for the reduction of the electrolyte redox couple (*i.e.*, $I_3^-/I^-$ and $Co^{+2/+3}$). The as-produced counter electrodes are tested in both $I_3^-/I^-$ and $Co^{+2/+3}$-based semitransparent DSSCs, showing power conversion efficiencies of 2.1% and 5.09%, respectively, under 1 SUN illumination. At 0.1 SUN, $Co^{+2/+3}$-based DSSCs achieve a power conversion efficiency as high as 6.87%. Our results demonstrate that the electrical, optical, chemical and catalytic properties of graphene-based dual films, designed by combining CVD-graphene and FLG flakes, are effective alternatives to FTO/Pt counter electrodes for DSSCs for both outdoor and indoor applications.

# 1. Introduction

Dye-sensitized solar cells (DSSCs)[1,2] represent an affordable photovoltaic (PV) technology (production costs below 0.5 USD/$W_p$[3]), which recently achieved power conversion efficiency (PCE) > 14%[4] (certified PCE of 14.1%[5]). These solar cells also offer several advantageous properties[6,7], such as lightweight[8], flexibility[9–11], low toxicity[12–14], and the possibility to efficiently work under weak and indirect illumination[15–17], (*e.g.*, in cloudy[18,19] and artificial light conditions)[20,21]. The DSSCs can be fabricated with on-demand color[22–25] and transparency level [26], and be thus integrated in buildings[27] or used for indoor energy generation[28–30]. Typically, a DSSC has a layered structure[2,31] composed of a conductive photoanode coated with a few-μm-thick semiconductor (*e.g.*, $TiO_2$[12,32,33], $ZnO$[34–36], $SnO_2$[37,38], $In_2O_3$[34,39], $Nb_2O_5$[40,41], and $SrTiO_3$[40]), a dye sensitizer (organic[14,42–46] or inorganic[12,47], *e.g.*, ruthenium bipyridyl derivatives[48,49] such as N719[49] and N3[49]), an electrolyte (liquid[50,51] or solid[50,52]) containing a redox system (*e.g.*, $I_3^-/I^-$ and $Co^{2+}/Co^{3+}$-complexes[53,54]) and a counter-electrode (CE) (*e.g.*, platinized transparent conductive oxide –TCO–[55–57] and nanocarbon-based electrodes[58–65]). More in details, CEs are composed by two components, namely the current collector (*e.g.*, TCO[55–57] or metal grid electrode[66,67]) and a catalytic film made of Pt particles[55–57] or nanocarbons[58–65]). The current collector gathers the electrons through the external circuit, while the catalysts regenerates the electrolyte species[2,31]. In most cases, fluorine-doped tin oxide (FTO) is used as current collector due to a high electrical conductivity (*i.e.*, sheet resistance –$R_{sheet}$– ≤ 10 Ω□$^{-1}$[68,69]) combined with optical transparency (transmittance –$T_r$– > 80% in the visible wavelength range[69,70]). Traditionally, platinized-FTO (FTO/Pt) have been adopted to catalyze the regeneration of the electrolyte species for the standard $I_3^-/I^-$ redox couple (*i.e.,* $I_3^- + 2e^- \rightarrow 3I^-$)[2,31]. Recently, both platinized-FTO (FTO/Pt)[71] or graphene nanoplatelets-coated FTO[46,72] have been exploited to reduce Co-based redox mediator[55,56]. Noteworthy, polypyridine complexes of $Co^{2+}/Co^{3+}$ (coupled with donor-π-bridge-acceptor sensitizers)[73,74] show redox potential (> 0.4 V *vs.* NHE[75,76]) more positive than the $I_3^-/I^-$ couple (0.35 V *vs.* RHE[76–80]), allowing PCE over 12% to be achieved[4,46,71,72] (up to 14.3%[4]). However, such results have been obtained by using current collectors of FTO, which are brittle[81] and whose fabrication requires high-temperature processes[82–84]. These undesired properties arise severe challenges for the FTO deposition on plastic substrates, as needed for flexible solar cells[85,86]. In addition, Pt-based CE can account for the 50% of the overall cell price [3], due to the high cost of Pt (~800 USD oz$^{-1}$)[87]. Pt-based CEs also tend to degrade when exposed to the liquid electrolytes [88,89], affecting the lifetime stability of the cells [88,89]. In this context, carbon-based materials have been successfully demonstrated as reliable alternative catalyst to replace Pt in CE. The list includes activated carbons[90], carbon nanotubes (CNTs) [91,92], hard carbon spherules[61], graphite[93–95], graphene flakes[96], graphene nanoplatelets[46,72,97], graphene oxide (GO)[98] and hybrid carbon nanocomposites[99,100]. Specially, graphene nanoplatelet deposited onto FTO have been reported as excellent catalyst for DSSC based on the Co(2,2′-bipyridine)$_3$$^{3+/2+}$ redox couple[46,101], outperforming FTO/Pt CE[101] and reaching PCE as high as 13%[46]. However, to date the development of (semi)transparent CEs entirely based on nanocarbons acting both as current collector and catalysts has been scarcely investigated[64,102,103]. For example, a CE based on CNT fibers has been recently proposed as dual-functional current-collectors/catalysts in DSSCs with PCE up to 8.8%[64]. However, the bottleneck regarding the use of "*bulky*" CNTs fibers is the loss of transparency, which limits the possibility of illuminating the DSSCs from both sides in a conventional configuration[64], as well as the fabrication of DSSCs with on-demand color[22–25] to be integrated in buildings[27].

In this context, graphene appears to be the ideal candidate to fulfill multiple functions[104,105] as it uniquely combines high charge carrier mobility (~200000 cm$^2$ V$^{-1}$ s$^{-1}$)[106,107] optical transparency[108] (transmittance –$T_r$– ~97.7%[109]), excellent thermal conductivity (~5000 W m$^{-1}$ K$^{-1}$) [110,111], outstanding mechanical properties[112] (*e.g.*, Young's modulus ~1 TPa[113]), and catalytic activity[46,72,96,97,114–118]. These properties refer to single graphene flakes and it is still difficult to match them in large-area samples; however, recent reports demonstrated the application of graphene-based electrodes in a wide variety of devices[119–123], such as solar cells[124–128], light emitting diodes[129–131], touch screens[132–134], field-effect transistors[135,136], photodetector[137] (electro)chemical sensors[138–140] and energy storage system[141–146].

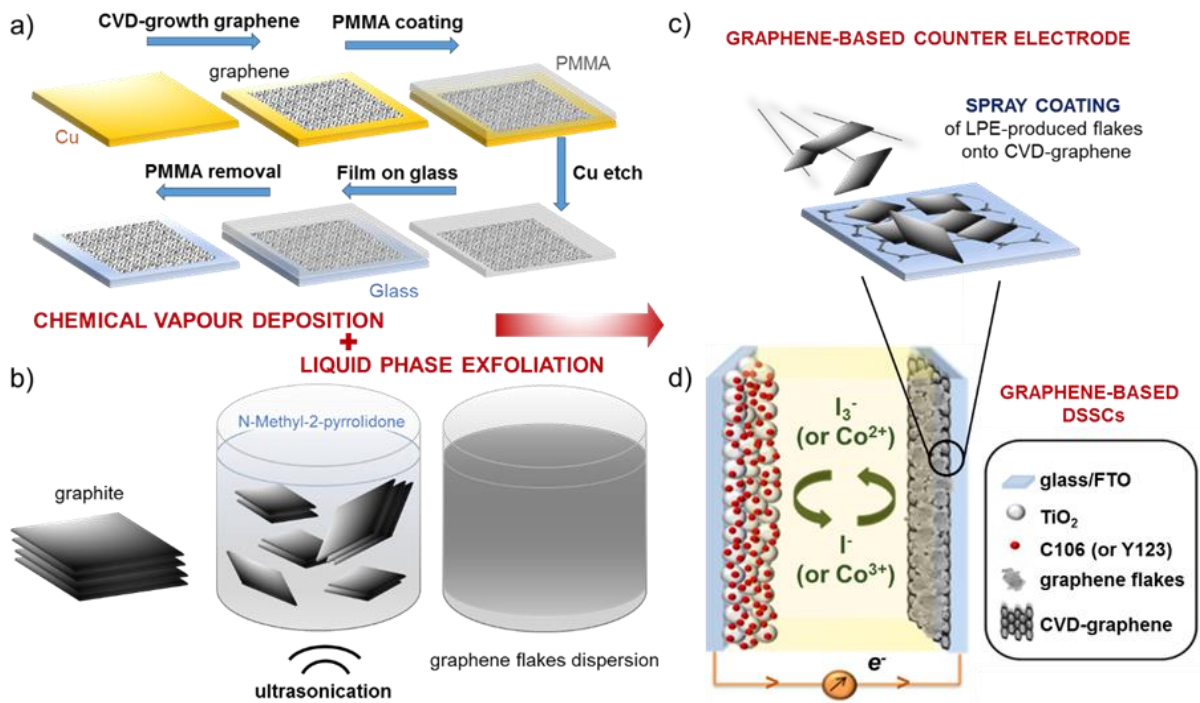

**Figure 1.** Sketch of the fabrication of DSSCs with CVD-graphene/graphene flakes dual films as CE. (a) Process flow of the CVD growth of graphene and its wet-transfer on glass substrate: a Cu foil was used as substrate for the growth; the Cu foil was dissolved by wet-etching; the graphene film was transferred by PMMA-assisted method onto a glass substrate; the PMMA layer was dissolved with acetone. (b) Production of graphene flakes dispersion through LPE of graphite. (c) Fabrication of CVD-graphene/graphene flakes dual films by spray coating graphene flakes over CVD-graphene. (d) Architecture of the DSSC using graphene-based CEs. Both $I_3^-/I^-$ and $Co^{+2/+3}$-based electrolyte have been tested, together with C106 and Y123 dye, respectively.

The success of such graphene-based electrodes, especially if compared to those based on others carbon allotropes, crucially originates from the implementation of effective production[147–149] and processing techniques[150–154]. In this context, large-area growth of pristine single-layer graphene (SLG) can be obtained *via* chemical vapor deposition (CVD) [155–158]. Subsequently, it can be transferred onto arbitrary substrates [159] by both wet [147,160–164] and dry [121,132,147,165] methods. This allows for the realization of flexible electrodes with $T_r > 85\%$ and $R_{sheet} \leq 30\ \Omega_\square^{-1}$ [132,166,167], which are both comparable to those of conventional TCO[168,169], *e.g.*, indium tin oxide –ITO–[170,171] and FTO[69,172]). Meanwhile, the production of graphene-based materials in high volumes have been achieved through liquid-phase exfoliation (LPE) methods[147,148,173–176]. These methods are based on top-down approaches which permit to exfoliate bulk graphite into SLG and few-layer graphene (FLG) in a liquid environment[147,148,173–176]. This process is performed by exploiting techniques such as ultrasonication[177–181], ball milling[182,183] shear mixing[184–187] and wet-jet milling[175,188] to break the van der Waals bond between the adjacent planes of the graphite. The possibility to produce and process graphene in a liquid phase allows functional inks with on-demand rheological and morphological properties to be formulated[173–175,189]. This represents a step forward towards the development of industrial-scale, reliable and inexpensive printing/coating processes of graphene[114,189–194].

Based on the advantages offered by the aforementioned graphene production and processing methods, we demonstrated a novel TCO and noble metal-free CE based on CVD-graphene/graphene flakes dual-film (**Figure 1**). By taking advantages of different graphene production and processing methods, we successfully eliminated the drawbacks typically related to the use of (i) Pt as catalyst and (ii) FTO as current collector. Specifically, low-resistance graphene film produced by CVD (Figure 1a) acts as charge collector, while high-surface area graphene flakes with abundant sharp edges, produced by liquid phase exfoliation (LPE) (Figure 1b) and spray-coated onto CVD-graphene (Figure1c), work as catalysts for the reduction of the electrolyte redox couple (Figure 1d). Chemical vapor-deposited graphene exhibited a high optical transparency ($T_r$ > 97% for visible wavelength ranging between 500-800 nm) and low $R_s$ (~1.2 kΩ□$^{-1}$), while submicron-sized graphene flakes provided high specific surface area –SSA– (123 ± 25 m$^2$g$^{-1}$) and sub-μm size (lateral dimension ranging between 200-900 nm), resulting in abundance of catalytic edge sites. The as-produced CE were tested in both $I_3^-/I^-$ and $Co^{+2/+3}$-based semi-transparent ($T_r$ of 13.8% at 700 nm) DSSCs (Figure 1d). The DSSCs using graphene-based counter electrode, commercial $I_3^-/I^-$ electrolyte, ruthenium complex dye (C106) have shown a PCE of 2.1% at AM 1.5 G illumination (1 SUN) (2.8% at 0.1 SUN). By using [Co(bpy-pz)$_2$]$^{2+/3+}$ (bpy-pz = 6-(1H-pyrazol-1-yl)-2,2'-bipyridine) as redox couple, and cyclopentadithiophene-bridged donor-acceptor dye (Y123) as organic dye[71], the graphene based DSSCs attained a power conversion efficiency (PCE) of 5.09% at 1 SUN (6.87% at 0.1 SUN). These results, coupled with the recent progress on designing metal-free organic sensitizer[14,42–45], quantum-dot sensitizer[195,196] perovskite-based sensitizer[197–199], and natural dyes,[14,45], make DSSCs to be an important PV technology.

## 2. Methods

### 2.1 Production of materials

*Chemical vapor deposition and transfer of graphene.* Continuous films of graphene were synthesized on Cu foil (Sigma Aldrich, thickness 20 μm, 99.999%) by CVD by using a cold-wall reactor with methane as a carbon precursor[200]. The Cu foil was loaded into the CVD reactor, which was then heated to 1060°C in Ar atmosphere to anneal the foil for 10 min. After this annealing step, the graphene growth was performed at 1060 °C for 10 min (pressure of 25 mbar) by flowing CH$_4$ (2 sccm), H$_2$ (20 sccm) and Ar (1000 sccm). The CVD reactor was then cooled down to 120 °C before removing the samples to prevent substrate oxidation. Samples of CVD-graphene of 1×1 cm$^2$ size were transferred to glass or SiO$_2$ substrates using wet transfer technique with poly(methyl methacrylate) (PMMA) as support medium[201]. Briefly, a thin layer of PMMA (2% solution in acetyl lactate, All resist GmbH) was deposited onto Cu/graphene by spin coating, and dried for 1 h at ambient conditions. The as-obtained samples were immersed in a 0.05 M solution of iron(III) chloride (FeCl$_3$, Sigma-Aldrich) for 16 h to etch the Cu and release the graphene/PMMA film. Once the Cu was completely etched away, the graphene/PMMA membrane was removed from the FeCl$_3$ solution using a glass slide and transferred to deionized water several times to remove the etchant residue. Subsequently, the membrane was removed from the water using glass or SiO$_2$ substrates, and dried at ambient conditions. Finally, the PMMA support film was removed by immersing the sample in acetone (Sigma Aldrich) for 4 h and then rinsed in 2-propanol (Sigma Aldrich).

*Graphene flakes production via LPE of graphite.* The graphene flakes were produced by the LPE[114,147,148,192], followed by SBS[114,189,192] of pristine graphite (+100 mesh, ≥ 75% min, Sigma Aldrich) in N-Methyl-2-pyrrolidone (NMP) (99.5% purity, Sigma Aldrich). Experimentally, 1 g of graphite was dispersed in 100 mL of NMP and ultrasonicated in a bath sonicator (Branson 5800 cleaner, Branson Ultrasonics) for 3 h. The resulting dispersion was then ultracentrifuged at 4300 g (in Beckman Coulter Optima XE-90 with a SW32Ti rotor) for 30 min at 15 °C, exploiting SBS to remove thick flakes and un-exfoliated graphite[202,203]. Subsequently, the supernatant (~80% of the dispersion) was collected by pipetting,

### 2.2 Characterization of materials

Scanning electron microscopy (SEM) images of CVD-graphene were taken with a FE-SEM (Jeol JSM-7500 FA). The acceleration voltage was set to 5 kV.

Transmittance spectra of the CVD-graphene on glass were taken with a Cary Varian 6000i UV*vis*-NIR spectrometer, using a 1 mm pinhole holder. The pristine glass substrate was used as baseline. Each sample was measured 5 times and the averaged values were reported.

Optical absorption spectroscopy (OAS) measurements of the LPE-produced graphene flake dispersion in NMP were performed with a Cary Varian 6000i UV*vis*-NIR spectrometer. The absorption spectra were acquired using a 1 mL quartz glass cuvette. The inks were diluted to 1:100 in NMP, to avoid scattering losses at higher concentrations. The corresponding solvent baseline was subtracted to each spectrum. The concentration of graphitic flakes is determined from the optical absorption coefficient at 660 nm, using $A = αlc$ where l is the light path length, c is the concentration of dispersed graphitic material, and α is the absorption coefficient, with α ~ 1390 L g$^{-1}$m$^{-1}$ at 660 nm[180,204].

Raman spectroscopy measurements on CVD-graphene (transferred on glass and Si/SiO$_2$) and LPE-produced graphene flakes were carried out by using a Renishaw inVia confocal Raman microscope using an excitation line of 532 nm (2.33 eV) with a 50× objective lens, and an incident power of ~ 1 mW on the samples. The LPE-produced flakes were obtained by drop-casting their dispersion in NMP onto a Si wafer with 300 nm of thermally grown $SiO_2$ (LDB Technologies Ltd.). The samples were then dried under vacuum before the measurement. The spectra were fitted with Lorentzian functions. Statistical analysis was carried out by means of OriginPro 9.1 software.

Transmission electron microscopy (TEM) measurements of the LPE-produced graphene flakes were carried out with a JEM 1011 (JEOL) transmission electron microscope operating at an acceleration voltage of 100 kV. The samples were obtained by depositing 1:100-diluted graphene flake dispersion in NMP onto holey carbon (200 mesh grids). Subsequently, the samples were dried under vacuum overnight.

Atomic force microscopy (AFM) images of the LPE-produced flakes were taken using an Innova AFM (Bruker, Santa Barbara, CA). The measurements were taken in tapping mode with a NTESPA 3.75 mm cantilever (Bruker, 300 kHz k: 40 N m$^{-1}$), in air at room temperature, with a relative humidity less than 30%. The software used for image analysis was Gwyddion version 2.43. Statistical analysis was carried out by means of Origin 9.1 software, using different AFM images of the sample. The sample for the measurements was prepared by drop-casting 1:30 diluted graphene flake dispersion in NMP onto mica sheets (G250-1, Agar Scientific Ltd., Essex, U.K.) and drying them under vacuum overnight.

### 2.3 Fabrication of graphene-based CEs

Glass/CVD-graphene were produced according to the protocols described in Section 2.1. Subsequently, the LPE-produced graphene flakes dispersions (see Section 2.1 for the details regarding the dispersion production) were deposited onto glass/CVD-graphene samples by spray coating. The dispersions were sprayed using a flux of $N_2$ at 1 bar, and keeping the substrates at a temperature of 150 °C. By controlling the amount of sprayed dispersion, three different mass loadings of the graphene flakes (0.16, 0.32 and 0.48 mg cm$^{-2}$) were deposited onto the CVD-graphene. After the deposition of the graphene flakes film, the substrates were annealed in glove box at 350°C for 2.5 h in order to remove solvent residuals.

### 2.4 Characterization of graphene-based CEs

Scanning electron microscopy (SEM) images were acquired with a FE-SEM (Jeol JSM-7500 FA). The acceleration voltage was set to 5 kV.

Atomic force microscopy and Raman spectroscopy measurements were acquired with the same instrumentations and parameters reported in Section 2.2.

Transmittance spectroscopy measurements of the graphene-based CEs were acquired with an integrating sphere-supported UV-Vis 2550 Shimadzu Spectrophotometer. A glass/CVD-graphene substrate was used as baseline. Each sample was measured 5 times and the average values were reported.

Sheet resistance measurements were performed with a Keithley Model 2612A Dual-channel System Source Meter in four-point probe configuration, using in line gold-plated probes of constant spacing (2 mm) contacting the surface of the films.

Specific surface area measurements of electrodes were carried out in Autosorb-iQ (Quantachrome) by Kr physisorption at temperatures of 77 K. The specific surface areas were calculated using the multipoint Brunauer-EmmettTeller (BET) model[205], considering equally spaced points in the $P/P_0$ range from 0.009 – 0.075 to. $P_0$ is the vapour pressure of Kr at 77 K, corresponding to 2.63 Torr[206–209]. Before the measurements, the samples were degassed for 1 h at 60 °C under vacuum conditions to eliminate weakly adsorbed species.

### 2.5 Fabrication of DSSCs and symmetrical dummy cells

Fluorine-doped $SnO_2$-coated glass (soda lime) substrates (8 Ω□$^{-1}$, Pilkington Tec) were cleaned using successive ultrasonic baths in acetone (Sigma Aldrich) and ethanol (Sigma Aldrich). Two types of DSSCs using C106 and Y123 as dye sensitizer, respectively, were fabricated using specific protocols. For C106-based DSSCs, films of nanocrystalline $TiO_2$ (0.5×0.5 cm$^2$) were deposited onto FTO-glass *via* screen-printing, using a $TiO_2$ paste (Dyesol 18NR-T). Subsequently, the substrates were dried in an oven for 20 min at 120 °C in order to evaporate the solvent. The thickness of $TiO_2$ layers was 6 μm thick, as measured by Dektak Veeco 150 profilometer. The as-obtained samples were then exposed to a sintering procedure at 525 °C for 30 min. Then, the samples were soaked in the C106 solutions for 16 h, washed with ethanol and blown with compressed air, obtaining the photoanodes. Graphene-based CEs were prepared as described in Section 2.3. Reference counter electrodes based on FTO-Pt were also fabricated. Briefly, a Pt layer was deposited onto FTO by screen-printing an as-purchased paste containing Pt precursor (Platisol, Solaronix). Subsequently, the substrates were dried in an oven at 120 °C for 20 min, and sintered for at 480 °C 30 min. The devices were laminated with a 25 μm-thick thermoplastic resin (Surlyn, Solaronix). After a hot-melting step, the distance between the two electrodes was measured to be about 20 μm.

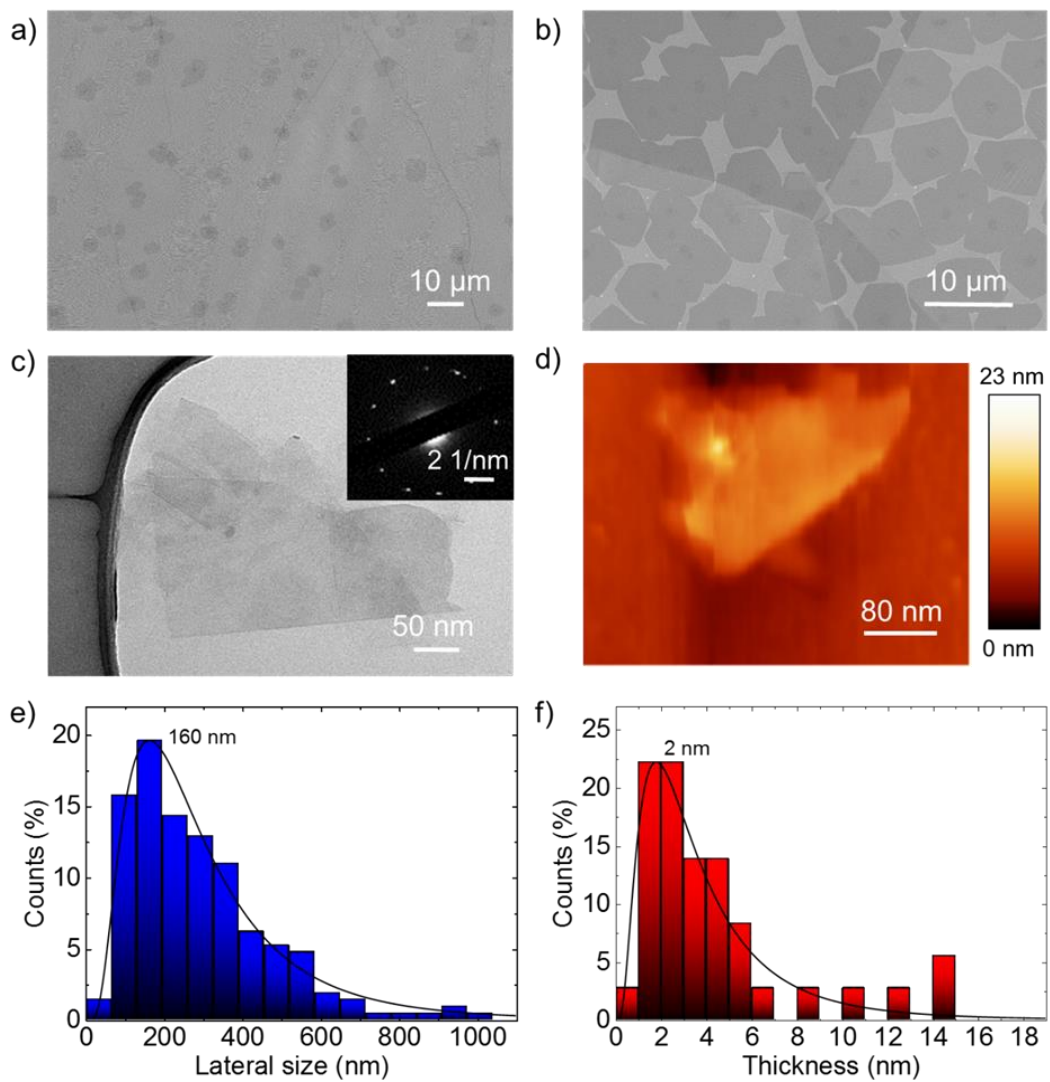

**Figure 2 Morphological characterization of CVD-graphene and LPE-produced graphene flakes.** (a) Representative SEM image of CVD-graphene polycrystalline film grown on Cu foil by using a cold-wall reactor and a growth time of 10 min. (b) SEM image of CVD-graphene obtained with a growth time of 5 min for estimating nucleation density and average grain size of the polycrystalline film. (c) Bright field TEM and (d) AFM images of LPE-produced graphene flakes. (e) Lateral size and (f) thickness statistical analyses of the LPE-produced graphene flakes (acquired on 80 flakes).

A commercial $I_3^-/I^-$-based electrolyte (High Performance Electrolyte –HPE–, Dyesol) was injected by vacuum assistance. Lastly, the cells were closed with glue. In the case of DSSC based on Y123 conductive Pilkington TEC glassy plates (4×15 cm$^2$) were immersed in 100 ml of TiCl$_4$/water solution (40 mM) at 70 °C for 30 min, washed with water and ethanol and dried in an oven at 80 °C for 30 min. The TiO$_2$ layers were deposited on the FTO glassy plates by screen-printing (frame with polyester fibers having 77.48 mesh cm$^{-2}$). This procedure, involving two steps (coating and drying at 125 °C), was repeated twice. The TiO$_2$-coated plates were gradually heated up to 325 °C. Then, the temperature was increased to 375 °C in 5 min, and afterwards to 500 °C. The plates were sintered at this temperature for 30 min, finally cooled down to room temperature. After the TiO$_2$ film was treated with 40 mM TiCl$_4$ solution, following the procedure previously described above, rinsed with water and ethanol. Lastly, a coating of TiO$_2$ nanoparticles (150-200 nm in size, Dyesol) was deposited as scattering layer onto the samples by screen-printing and sintered at 500°C.

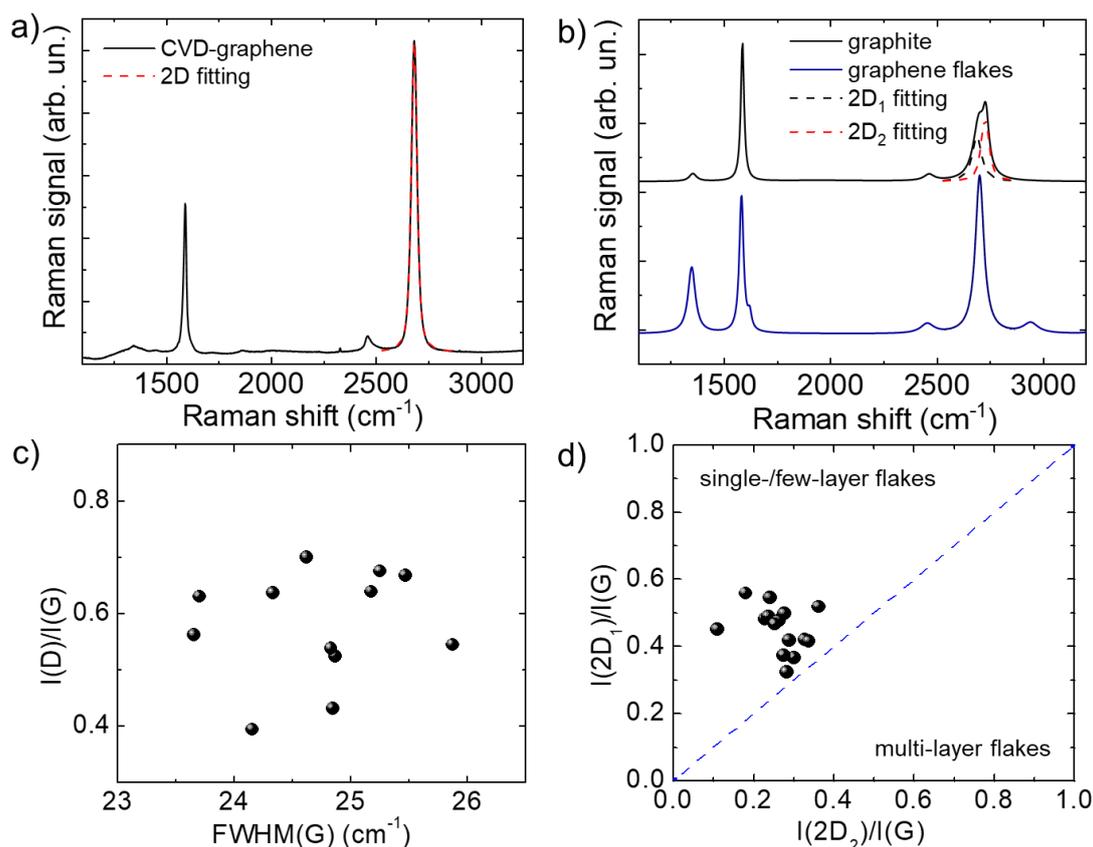

**Figure 3.** (a) Raman spectrum of CVD-graphene. (b) Comparison between the Raman spectra of the graphite and LPE-produced graphene flakes. The multi-peak Lorentzian fittings of the Raman spectra in panels (a) and (b) show the contribution of the different 2D modes, *i.e.*, $2D_1$, $2D_2$. (c) Plot of $I(D)/I(G)$ *vs.* FWHM(G). (d), Plot of $I(2D_1)/I(G)$ *vs.* $I(2D_2)/I(G)$. The dashed line $I(2D_1)/I(G) = I(2D_2)/I(G)$ represents the multi-layer condition (~5 layers).[210] The Raman statistical analysis was carried out on 15 spectra.

Each anode was cut into rectangular pieces (area: 2×1.5 cm$^2$) having a masked spot area of 0.181 cm$^2$ with a total thickness of ~8 μm. The anode were soaked for 16 h in a dye solution composed by Y123 (0.1 mM) and chenodeoxy acid (5mM) in alcohol ter-butylic:acetonitrile (1:1). The assembly of the whole DSSCs followed the same protocols adopted for C106-based DSSCs, except for the use of Co(bpy-pz)$_2^{(2+/3+)}$ (bpy-pz= 6-(1H-pyrazol-1-yl)-2,2'-bipyridine)-based electrolyte (DN C09 and DN C10, Dyenamo).

Symmetrical dummy cells were produced by assembling two identical CEs of FTO, or FTO/graphene flakes, or FTO/Pt, or CVD-graphene/graphene flakes. The electrodes were sealed with 25 μm-thick thermoplastic resin (Surlyn, Solaronix) filled with $I_3^-/I^-$ electrolyte (HSE, Dyesol). An active area of 1.44 cm$^2$ was made with a thermoplastic mask.

## 2.6 Characterization of DSSCs and dummy cells

The PV performance of DSSCs was determined by current density *vs.* voltage (IV) measurement in air under a solar simulator (ABET Sun 2000, class A) at 1 SUN and 0.1 SUN, calibrated with a certified reference Si Cell (RERA Solutions RR-1002). The incident power was measured with a Skye SKS 1110 sensor. The class of the sun simulator was measured with a BLACK-Comet UV-VIS Spectrometer.

Electrochemical impedance spectroscopy (EIS) measurements of dummy cells were taken in dark conditions at room temperature using an Autolab 302N Modular Potentiostat (Metrohm) in two-electrode configuration under short circuit condition (0 V of electrical bias). The ac perturbation was set at 10 mV with frequencies ranging from 100 kHz to 0.1 Hz.

## 2. Results and discussion

The CVD-graphene and the LPE-produced graphene flakes (see Experimental, Section 2.1, for the material production details) were first characterized separately before the fabrication of the graphene-based CE. The morphology of the materials was evaluated by electron microscopy and AFM measurements.

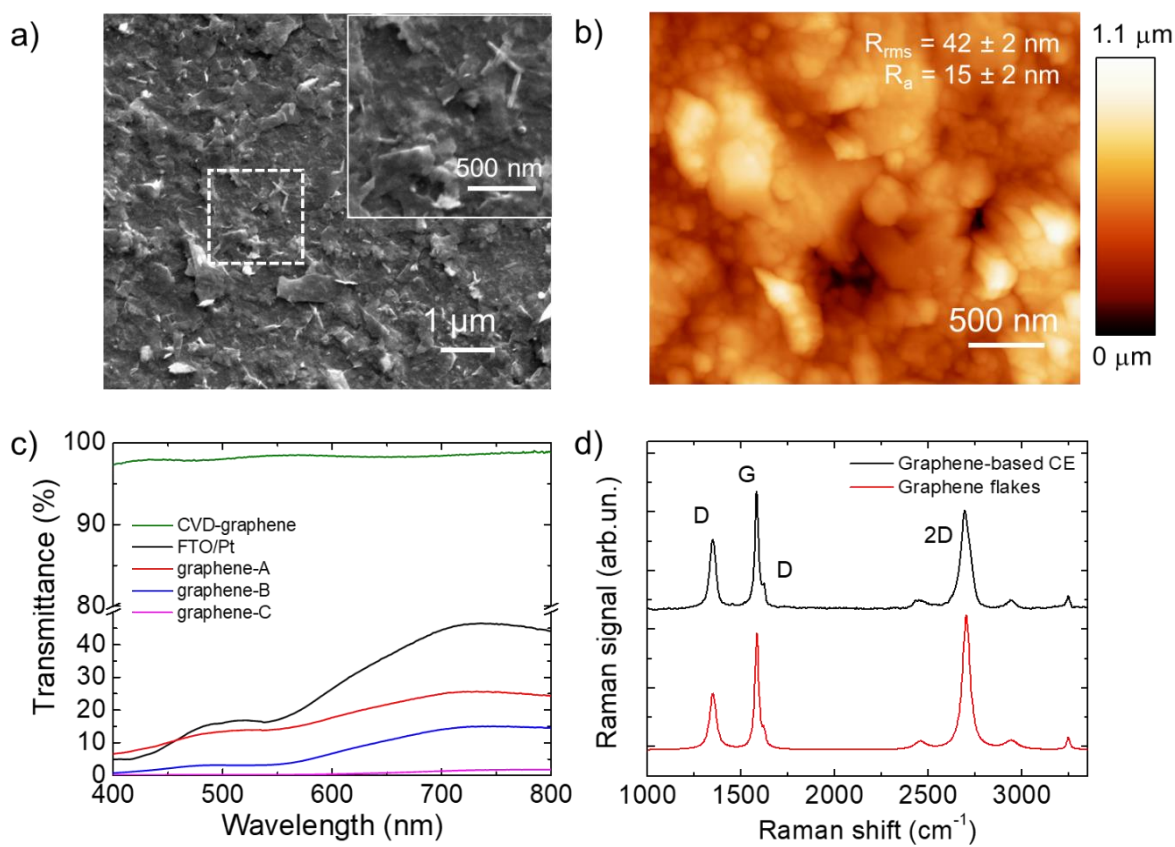

**Figure 4**. (a) SEM and (b) AFM images of the surface of a representative graphene-based CE (graphene-B). The inset to panel (a) shows a high-magnification SEM image of the area delineated by the dashed white line. The $R_{rms}$ and $R_a$ values are reported in panel (b). (c) Transmittance spectra of the graphene-based CEs. The transmittance spectra of a conventional FTO/Pt CE is also shown for comparison. (d) Comparison between the Raman spectra of LPE-produced graphene flakes (red) and a corresponding graphene-based CE (black) (with a mass loading of graphene flakes of 0.32 mg cm$^{-2}$).

**Figure 2**a shows a representative SEM image of the CVD-graphene grown on Cu foil by using a cold-wall reactor (see additional detail in Experimental, Section 2.1)[200]. The sample is overall a continuous polycrystalline film without any tears or holes, in agreement with previous studies[200]. Such morphology resulted in a $R_{sheet}$ of ~1.2 kΩ□$^{-1}$. Partial coverage experiments with a growth time of 5 min revealed a high nucleation density (25000 grains mm$^{-2}$) with an average grain size of a ~10 μm (Figure 2b). The nucleation of graphene on Cu occurs primarily on the surface irregularities such as the rolling grooves and crystal terraces acting as energetically favorable spots for the nucleation of graphene[200,211]. Figure 2c,d show representative bright field TEM and AFM images, respectively, of the LPE-produced sample, consisting of irregularly shaped (Figure 2c) and nm-thick flakes (Figure 2d). The corresponding electron diffraction pattern of the TEM image is also shown (inset in Figure 2c), proving that the flakes are crystalline[212]. Statistical analysis (Figure 2e,f) indicate that the lateral size and the thickness of the flakes follow a lognormal distribution, peaked at ~190 nm and ~2 nm, respectively. These data indicate that the sample obtained by LPE of graphite is mostly composed by FLG flakes (experimental SLG thickness measured by AFM is typically between 0.4 and 1 nm[104,213–215] depending on tip-surface interactions and image feedback settings[214,215]). Raman spectroscopy analysis was carried out to evaluate the structural properties of the as-produced materials. A typical Raman spectrum of graphene shows, as fingerprints, G, D and 2D peaks (see Supplementary material for more details)[192,216–218]. For SLG the 2D band is roughly four times more intense than the G peak[216,217]. Multi-layer graphene (> 5 layers) exhibits a 2D peak, which is almost identical, in term of intensity and lineshape, to the graphite case

(intensity of the 2D$_2$ band is twice the 2D$_1$ band)[217–219]. Instead, FLG (< 5 layers), has a 2D$_1$ peak more intense than the 2D$_2$[220]. Taking into account the intensity ratios of the 2D$_1$ and 2D$_2$, it is possible to estimate the flake thickness[145,175,192,193,220]. **Figure 3**a shows the Raman spectrum of the CVD-graphene transferred onto Si/SiO$_2$. The absence of the defect-related D peak and the ratio between the intensity of 2D and IG peak –I(2D)/I(G)– of ~ 3 indicate that a high-quality SLG has been obtained[221], in agreement with previous studies[200]. The analysis of the 2D peak, a single and sharp Lorentzian band centered at ~2683 cm$^{-1}$, also confirm that the sample is SLG[192,216,218]. Figure 3b displays a representative Raman spectrum of LPE-produced graphene flakes. The Raman spectrum of the native bulk graphite is also shown for comparison. The LPE-produced flakes exhibit an enhancement of the D and D' bands compared to those of the pristine graphite, in agreement with previous discussion[192,216,218,222–225]. For graphite, the G peak is more intense of D, while the Raman statistical analysis for LPE-produced graphene flakes (**Figure S1**) shows that the ratio between the intensities of the D and G peaks –I(D)/I(G)– ranges between 0.35 and 0.75, with an average values of ~0.55 (Figure S1a). However, the plot of I(D)/I(G) *vs.* FWHM(G) (Figure 3c) does not show a linear correlation, which means that the defects mainly originate from the flake edges without altering the structure of the basal plane[226,227]. The analysis of I(2D$_1$)/I(2D$_2$) (Figure 3d) demonstrates that the LPE-produced sample has a few-layer flakes enriched composition[210,217,220].

After the preparation and characterization of the CVD-graphene and LPE-produced graphene flakes, the CEs were fabricated by spraying the LPE-produced graphene flakes dispersion onto the CVD-graphene previously transferred onto the glass substrate, as detailed in Section 2.3. The concentration of graphene flakes in the LPE-produced dispersion (~0.36 gL$^{-1}$) was estimated by OAS measurements (**Figure S2**) [204], and served to adjust the amount of deposited graphene flakes. Three different batch of graphene-based CE (named as graphene-A, graphene-B, graphene-C) were prepared with graphene flakes mass loading of 0.16, 0.32, 0.48 mg cm$^{-2}$, respectively. The morphology of the CEs was analyzed by SEM and AFM measurements. **Figure 4**a reports a SEM image of a representative graphene-based CE (graphene-B). The electrode surface is completely covered by a film of graphene flakes. Figure 4b shows the AFM images of the surface of the same electrode. The root means square roughness (R$_{rms}$) of the sample is 42 ± 2 nm and the average roughness (R$_a$) is 15 ± 2 nm. The SSA of the electrode was estimated by Brunauer, Emmett and Teller (BET) analysis [205] of physisorption measurement with Kr at 77 K [206,208,209], obtaining a value of ~123 ± 25 m$^2$ g$^{-1}$. Figure 4c shows the transmittance spectra of the CVD-graphene/FLG flakes dual film CEs, in comparison to those measured for both the CVD-graphene and conventional FTO/Pt CE (see fabrication detail in Experimental section). As previously reported, CVD-graphene shows an excellent optical transparency (T$_r$ > 97% for visible wavelengths ranging between 400 and 800 nm). After graphene flakes deposition, graphene-A and graphene-B retain T$_r$ ~25% and ~14%, respectively, at 700 nm, while graphene-C exhibit low T$_r$ (< 5% for all the visible wavelength) due to the high mass loading of the graphene flakes (*i.e.*, 0.48 mg cm$^{-2}$). Figure 4d reports the Raman spectroscopy analysis of the graphene-based CE. The Raman spectrum of the graphene-based electrode resembles those obtained for the LPE-produced flakes, thus indicating that the spray deposition process did not affect the structural properties of the flakes.

The graphene-based CEs were used to fabricate DSSCs by using commercial I$_3^-$/I$^-$-based electrolyte, and C106 dye (see detail of the DSSCs' fabrication in the Experimental section), which are simply named the corresponding CEs (*i.e.*, graphene-A, graphene-B and graphene-C). **Figure 5**a shows the IV measurements of the devices under 1 SUN and in dark (inset to panel), respectively. **Table 1** summarizes the main PV parameters of the fabricated DSSCs (as estimated by the IV curves). DSSCs with CE based on FTO and CVD-graphene are also shown for comparison. The PCE of such cells is less than 0.01%, indicating the need of a catalytic layer for obtaining an efficient electrolyte regeneration process. It is worth noticing that DSSCs based on only graphene flakes have been preliminary discarded because of the high R$_{sheet}$ of their films (> 100 kΩ□$^{-1}$ for a mass loading of 0.16 mg cm$^{-2}$ and > 10 kΩ□$^{-1}$ for both mass loadings of 0.32, 0.48 mg cm$^{-2}$), which resulted in poor PV performances. Graphene-A exhibits a short circuit current –J$_{sc}$– of 5.67 mA cm$^{-2}$, an open circuit voltage –V$_{oc}$– of 740 mV, and a fill factor –FF– of 37.2%, leading to a PCE of 1.62%. By increasing the mass loading of the graphene flakes, the PCE of graphene-B (2.13%) improves by 51% compared to graphene-A. This is mainly attributed to the higher J$_{sc}$ and FF of graphene B (5.97 mA cm$^{-2}$ and 54.9%, respectively) compared to those of graphene-A. However, a further increase of the mass loading of graphene flakes (graphene-C) decreases the PCE to 1.2%. The IV results can be tentatively explained by considering that the catalytic activity depends both by the amount of the catalytic materials and its electrical resistance. In absence of resistive effects, the increase of the mass loading of the catalytic materials enhances the catalytic activity of the CEs for the electrolyte regeneration reaction ($I_3^-$ reduction in our case), as typically observed for conventional FTO/Pt CE[228,229]. However, an excessive mass loading of graphene flakes can increase the electrical resistance of the corresponding layer, negatively affecting the catalytic activity of the superficial (*i.e.*, more exposed) flakes, which interact more effectively with the electrolytes than the inner flakes. The IV curves in dark (inset in Figure 5a) show that current density at forward bias polarity for graphene-A and graphene-C (maximum current density of 2.1 mAcm$^{-2}$ and 1.7 mAcm$^{-2}$, respectively) is lower than that of graphene-B (maximum current density of 12.0 mA cm$^{-2}$). Differently from the reference cell adopting FTO/Pt CE, the current density for the graphene-based DSSCs does not saturate, which means that the diffusion limiting current density is not reached in the investigated voltage window (where corrosion effects can be excluded)[230], and resistive losses occur.

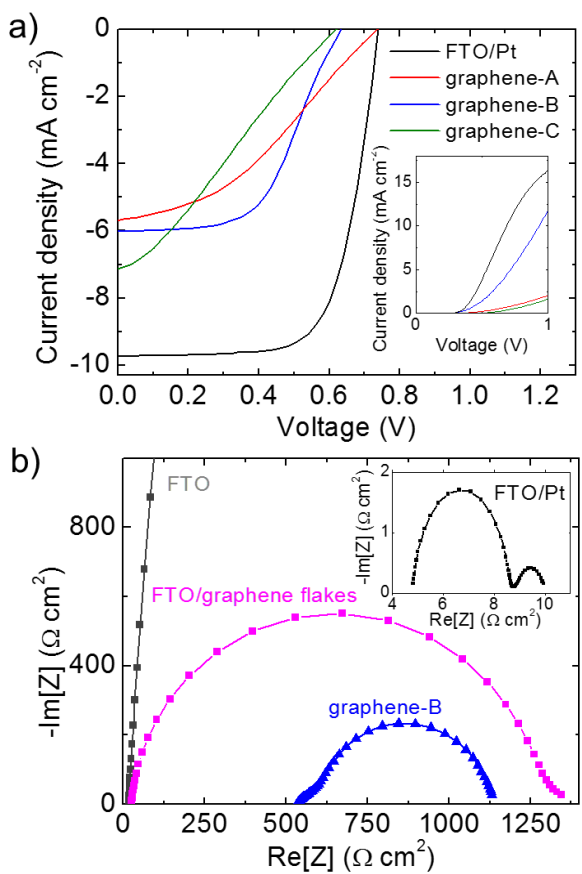

**Figure 5.** (a) IV curves of the DSSC based on C106 dye adopting $I_3^-/I^-$-based electrolyte and using graphene-based CE (graphene-A, graphene-B, graphene-C), compared to that of FTO/Pt CE-based reference, at 1 SUN. The inset shows the IV curve of the DSSCs obtained in dark. (b) Nyquist plots obtained by EIS measurements of symmetrical dummy cells based on FTO, FTO/graphene flakes, graphene-based CE (graphene-B) and FTO/Pt CE.

**Table 1.** Photovoltaic parameters of the DSSCs based on C106 dye and $I_3^-/I^-$-based electrolyte (at 1 SUN).

| Sample | Jsc (mA cm$^{-2}$) | Voc (V) | FF (%) | PCE (%) |
|---|---|---|---|---|
| FTO[*] | 0.05 | 0.55 | 10.6 | <0.01 |
| CVD-Graphene[*] | 0.04 | 0.77 | 23.9 | <0.01 |
| FTO/Pt | 9.72 | 0.75 | 68.7 | 4.9 |
| graphene-A | 5.67 | 0.74 | 37.2 | 1.6 |
| graphene-B | 5.97 | 0.63 | 54.9 | 2.1 |
| graphene-C | 7.11 | 0.62 | 27.1 | 1.2 |

[*]Comparative DSSCs without catalytic active films (*i.e.*, Pt or graphene flakes).

In order to elucidate the origin of such losses in presence of graphene-based CEs, EIS measurements were carried out on dummy cells consisting of two identical CEs sandwiching the $I_3^-/I^-$-based electrolyte used for DSSCs (see detail of dummy cell fabrication in Experimental, Section 2.5). Electrochemical impedance spectroscopy is a perturbative techniques, which measures the current response of an electrochemical system when AC voltages are applied with difference frequency, thus computing the electrochemical impedance (Z)[231]. The analysis of Z allows the kinetics of the (photo)electrochemical processes[231], including the electronic and ionic ones occurring in the DSSCs, to be studied[230,232–234]. In particular, by performing EIS on symmetrical dummy cells it is possible to elucidate both the electrochemical activity and the electrical resistance of CEs in simulated DSSC operating conditions, eliminating the photoanode contribution[101,235,236]. The Z data can be expressed graphically in a Nyquist plot (imaginary part of Z –Im[Z]– *vs.* real part of Z –Re[Z]–), which is typically composed of two semicircles for symmetrical dummy cells[232,237]: the first one (at frequency in the kHz-MHz range) is associated to the catalyst/electrolyte interface[232,237], the second one (at lower frequency, < 100 Hz) is connected with ionic diffusion processes occurring in the electrolyte[232,237].

**Table 2.** Photovoltaic parameters of our DSSCs (data extrapolated by IV curves in Figure 6) in comparison to literature data about cells based on Y123 dye and Co(bpy-pz)2(2+/3+) (bpy-pz= 6-(1H-pyrazol-1-yl)-2,2'-bipyridine)-based electrolyte. Our DSSCs with all-graphene CEs were fabricated using CVD-graphene and graphene flakes (graphene-B), as described in the text. PEDOT:PSS stands for poly(3,4-ethylenedioxythiophene)-poly(styrenesulfonate), GNPs for graphene nanoplatelets, and PProDOT for poly(3,4-propylenedioxythiophene).

| CE | $J_{sc}$ (mA cm$^{-2}$) | $V_{oc}$ (V) | FF (%) | PCE (%) | Illumination intensity (SUN) | Ref. |
|---|---|---|---|---|---|---|
| FTO/Pt | 13.68 | 1.010 | 57 | 7.96 | 1 | this work |
| CVD-graphene/graphene flakes (graphene-B) | 11.23 | 0.958 | 47 | 5.09 | 1 | this work |
| FTO/Pt | 13.40 | 0.802 | 63 | 6.90 | 1 | [244] |
| Ag/PEDOT:PSS | 13.40 | 0.790 | 66 | 7.00 | 1 | [244] |
| FTO/Pt | 12.10 | 1.047 | 63 | 8.10 | 1 | [238] |
| FTO/GNPs | 12.70 | 1.030 | 70 | 9.30 | 1 | [238] |
| FTO/Pt | 13.45 | 1.015 | 69 | 9.52 | 1 | [243] |
| FTO/PProDOT | 13.06 | 0.998 | 77 | 10.08 | 1 | [243] |
| FTO/Pt | 1.47 | 0.930 | 66 | 9.00 | 0.1 | this work |
| CVD-graphene/graphene flakes (graphene-B) | 1.28 | 0.865 | 62 | 6.87 | 0.1 | this work |
| FTO/Pt | 1.31 | 0.934 | 77 | 9.93 | 0.1 | [243] |
| FTO/PProDOT | 1.35 | 0.914 | 81 | 10.56 | 0.1 | [243] |
| FTO/Pt | 1.37 | 0.984 | 74 | 10.00 | 0.1 | [238] |
| FTO/GNPs | 1.34 | 0.956 | 76 | 9.70 | 0.1 | [238] |

The intercept with the x-axis of the first semicircle represents the ohmic series resistance ($R_s$) given by both the electrolyte resistance and the resistance of external circuit, including the electrical resistance of the CEs[232,237–240]. The diameter of the first semicircle is proportional to the electrode-electrolyte charge transfer resistance ($R_{ct}$)[232,237–240], which is related to catalytic activity of the CE towards the electrolyte redox reactions (low $R_{ct}$ corresponds to an high catalytic activity)[96,231]. **Figure S3** reports the equivalent electrical circuit used to model the symmetrical dummy cells impedance, together with its description, in agreement with previous literature[231,240]. Figure 5b reports the Nyquist plots obtained for the symmetric dummy cells adopting different electrodes, *i.e.*, graphene-B and FTO/Pt. Fluorine-doped tin oxide and graphene flakes-covered FTO (FTO/graphene flakes) were also characterized to ascertain the impedance contribution of each layer, *i.e.*, the current collectors (FTO or CVD-graphene) and the catalytic films (Pt or graphene flakes). The FTO-based dummy cell has a low $R_s$ ~20 Ωcm$^2$, mainly ascribable to the $R_{sheet}$ of FTO (8 Ω□$^{-1}$), and a high $R_{ct}$ (> 10 kΩcm$^2$), which indicates the absence of catalytic activity[241]. The FTO/graphene flakes-based dummy cell exhibits the same $R_s$ of pristine FTO, while $R_{ct}$ (~650 Ωcm$^2$) decreases by one order of magnitude compared to that of FTO. For graphene-B-based dummy cell, $R_s$ is higher than 500 Ωcm$^2$ due to high $R_{sheet}$ of CVD-graphene (~1.2 kΩ□$^{-1}$), as measured by four-point probe. Interestingly, the $R_{ct}$ of graphene-B-based dummy cell (~250 Ωcm$^2$) decrease by ~62% compared to that of FTO/graphene flakes-based dummy cell. This indicates that the coupling between CVD-graphene and graphene flakes is effective for increasing the overall CE catalytic activity of the same graphene flakes. However, the $R_{ct}$ measured for graphene-B-based dummy cell is still significantly higher than that of FTO/Pt (< 2 Ωcm$^2$), which indicate that graphene-based CEs have an insufficient catalytic activity with $I_3^-/I^-$-based electrolyte for practical DSSCs operating under solar illumination, in agreement with previous studies[236,242].

In order to exploit CVD-graphene/graphene flakes as CE for efficient DSSC under simulated 1 SUN for outdoor application, we tested them into DSSCs based on Y123 dye and Co(bpy-pz)$_2^{(2+/3+)}$ (bpy-pz= 6-(1H-pyrazol-1-yl)-2,2'-bipyridine)-based electrolyte. It has been demonstrated that graphene-based materials exhibit a catalytic activity for reducing the redox mediator of polypyridine complexes of $Co^{2+}/Co^{3+}$ comparable to that of Pt[55,56,236,238,243].

**Figure 6**a shows the comparison between the IV curves obtained for a $Co^{2+}/Co^{3+}$-based DSSC adopting the optimized graphene-B (*i.e.*, the best graphene-based CE in $I_3^-/I^-$-based DSSCs) and FTO/Pt. In particular, the PCE of graphene-based DSSC achieved a PCE of 5.09%, which is much higher than the one reached with $I_3^-/I^-$-based electrolyte (2.1%). We further extended the study and validation of our graphene-based DSSCs for indoor applications, by testing them under low illumination intensity conditions (< 1 SUN). In this case, the photocurrent densities are lower than those generated under 1 SUN, and consequently the PV performances are less affected by the CE series resistance. Under 0.1 SUN illumination, the graphene-based DSSC reached a PCE of 6.87% (Figure 6b), *i.e.*, an increase of ~35% in comparison to the PCE measured at 1 SUN. This PCE increase at 0.1 SUN is mainly attributed to a higher FF (up to 0.62), which is instead limited at 1 SUN illumination (up to 0.47) due to the high $R_s$ of the CVD-graphene current collector.

**Table 2** summarizes the PV parameters of innovative DSSCs using Y123 dye and $Co^{2+}/Co^{3+}$-based electrolytes. At 1 SUN illumination, our graphene-based DSSCs have shown PV parameters approaching those of cells based on Ag/PEDOT:PSS CEs, showing an overall PCE of ~5.1% *vs* 7% [244]. The main difference here is related to the Ag electrode, which clearly provide lower $R_s$ than CVD-graphene, as demonstrated by the higher FF (66% vs 47%). This effect is further confirmed by the PV performances of the cells that used GNPs or PProDOT catalysts in combination with FTO. In these cases, the low $R_s$ of the TCO contributes in bringing the FF up to 70% and 77%, respectively, leading to PCE beyond 9%. Overall, our graphene-based CEs can be used in high-performance TCO/Pt-free DSSCs, which can compete with state-of-the-art DSSCs made with glass/FTO in combination with various kinds of carbon- and polymer-based catalysts[245]. This indicates the potential of the CVD-graphene/graphene flakes dual-films as CEs in DSSCs, providing a viable alternative to be experimented also in the context of flexible devices.

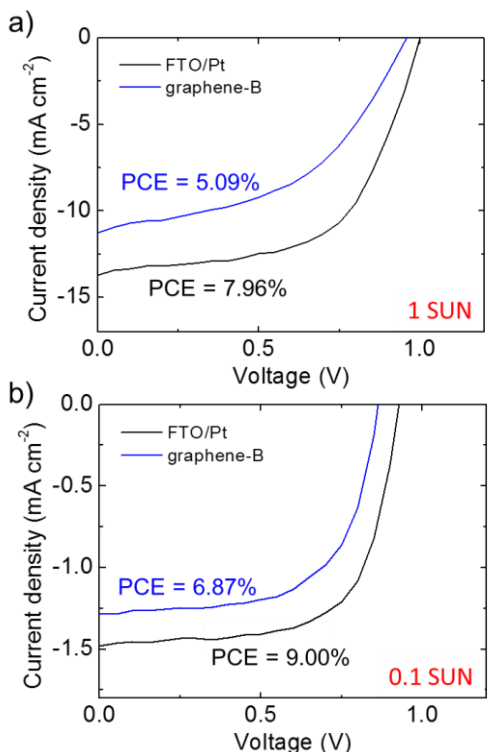

**Figure 6.** (a) IV curves of the DSSC based on Y123 dye adopting Co(bpy-pz)$_2^{(2+/3+)}$ (bpy-pz= 6-(1H-pyrazol-1-yl)-2,2'-bipyridine)-based electrolyte and using graphene-based CE (graphene-B) or conventional FTO/Pt CE, at 1 SUN. (b) IV curve of the same DSSCs reported in panel at low intensity illumination (0.1 SUN).

## 3. Conclusions

In this work, we demonstrated advanced Pt- and transparent conductive oxide (TCO)-free counter electrodes (CEs) for dye-sensitized solar cells (DSSCs) based on dual films of graphene materials. In particular, single-layer graphene, produced by chemical vapor deposition (CVD) using a cold-wall reactor, has been used as current collector, while graphene flakes produced by liquid phase exfoliation (LPE) effectively acted as catalyst for reducing the electrolyte. The graphene-based DSSCs have been tested in configuration using both $I_3^-/I^-$- and $Co^{2+}/Co^{3+}$-based electrolytes. In the first case, the optimized DSSCs using ruthenium complex dye (C106) reached a power conversion efficiency (PCE) of 2.1% at AM 1.5 G illumination (1 SUN. In the second case, [Co(bpy-pz)$_2$]$^{2+/3+}$ (bpy-pz = 6-(1H-pyrazol-1-yl)-2,2'-bipyridine) was selected as redox couple, and cyclopentadithiophene-bridged donor-acceptor dye (Y123) as organic dye [71]. The graphene-based DSSCs achieved a PCE of 5.09% at 1 SUN. Noteworthy, the DSSCs have shown even better performance under low illumination intensity condition, reaching a PCE of 6.87% at 0.1 SUN. To the best of our knowledge, these PCE values, compete with record-high values reported for DSSCs using Pt- and TCO-free CEs[238,243]. In perspective, the replacement of conventional FTO/Pt CE with all-graphene-based CEs is promising for lowering the manufacturing costs of DSSCs, as well as for developing flexible solar cells designs. Our results, coupled with the recent progress on the synthesis of metal-free organic sensitizer[14,42–45], quantum-dot sensitizer[195,196], perovskite-based sensitizer[197–199], and natural dyes[14,45], confirm DSSCs as an important branch within the current PV panorama.

## Acknowledgements

This project has received funding from the European Union's Horizon 2020 research and innovation program under grant agreement no. 696656–GrapheneCore1 and no.785219-GrapheneCore2. The authors also thank IIT Electron Microscopy for the access to carry out TEM and SEM measurements, and Alberto Portone for his help with transmittance measurements. A. Di Carlo gratefully acknowledges the financial support of the Ministry of Education and Science of the Russian Federation in the framework of Megagrant N°14.Y26.31.0027.

## References

[1] Grätzel M 2003 Dye-sensitized solar cells *J. Photochem. Photobiol. C Photochem. Rev.* **4** 145–53

[2] Hagfeldt A, Boschloo G, Sun L, Kloo L and Pettersson H 2010 Dye-Sensitized Solar Cells *Chem. Rev.* **110** 6595–663

[3] Kalowekamo J and Baker E 2009 Estimating the manufacturing cost of purely organic solar cells *Sol. Energy* **83** 1224–31

[4] Kakiage K, Aoyama Y, Yano T, Oya K, Fujisawa J and Hanaya M 2015 Highly-efficient dye-sensitized solar cells with collaborative sensitization by silyl-anchor and carboxy-anchor dyes *Chem. Commun.* **51** 15894–7

[5] NREL 2018 www.nrel.gov/pv/assets/images/efficiency-chart.png *accessed March 03, 2018*

[6] Hardin B E, Snaith H J and McGehee M D 2012 The renaissance of dye-sensitized solar cells *Nat. Photonics* **6** 162

[7] Albero J, Atienzar P, Corma A and Garcia H 2015 Efficiency Records in Mesoscopic Dye-Sensitized Solar Cells *Chem. Rec.* **15** 803–28

[8] Gong J, Liang J and Sumathy K 2012 Review on dye-sensitized solar cells (DSSCs): Fundamental concepts and novel materials *Renew. Sustain. Energy Rev.* **16** 5848–60

[9] Sipaut A R Y and R F M and C S 2015 Review on Metallic and Plastic Flexible Dye Sensitized Solar Cell *IOP Conf. Ser. Mater. Sci. Eng.* **78** 12003

[10] Weerasinghe H C, Huang F and Cheng Y-B 2013 Fabrication of flexible dye sensitized solar cells on plastic substrates *Nano Energy* **2** 174–89

[11] Yun M J, Cha S I, Seo S H and Lee D Y 2014 Highly Flexible Dye-sensitized Solar Cells Produced by Sewing Textile Electrodes on Cloth *Sci. Rep.* **4** 5322

[12] O'Regan B and Grätzel M 1991 A low-cost, high-efficiency solar cell based on dye-sensitized colloidal TiO2 films *Nature* **353** *737*

[13] Hug H, Bader M, Mair P and Glatzel T 2014 Biophotovoltaics: Natural pigments in dye-sensitized solar cells *Appl. Energy* **115** 216–25

[14] Calogero G, Bartolotta A, Di Marco G, Di Carlo A and Bonaccorso F 2015 Vegetable-based dye-sensitized solar cells *Chem. Soc. Rev.* **44** 3244–3294

[15] Tang Q, Wang J, He B and Yang P 2017 Can dye-sensitized solar cells generate electricity in the dark? *Nano Energy* **33** 266–71

[16] Rasheduzzaman M, Pillai P B, Mendoza A N C and Souza M M De 2016 A study of the performance of solar cells for indoor autonomous wireless sensors *2016 10th International Symposium on Communication Systems, Networks and Digital Signal Processing (CSNDSP)* pp 1–6

[17] Salvador P, Hidalgo M G, Zaban A and Bisquert J 2005 Illumination Intensity Dependence of the Photovoltage in Nanostructured TiO2 Dye-Sensitized Solar Cells *J. Phys. Chem. B* **109** 15915–26

[18] Berginc M, Kraŝovec U O and Topič M 2014 Outdoor ageing of the dye-sensitized solar cell under different operation regimes *Sol. Energy Mater. Sol. Cells* **120** 491–9

[19] Lepikko S, Miettunen K, Poskela A, Tiihonen A and Lund P D 2018 Testing dye-sensitized solar cells in harsh northern outdoor conditions *Energy Sci. Eng.* **6** 187–200

[20] Long Y, Hsu S and Wu T 2016 Energy harvesting characteristics of emerging PV for indoor and outdoor *2016 IEEE 43rd Photovoltaic Specialists Conference (PVSC)* pp 796–801

[21] Wu T-C, Long Y-S, Hsu S-T and Wang E-Y 2017 Efficiency Rating of Various PV Technologies under Different Indoor Lighting Conditions *Energy Procedia* **130** 66–71

[22] Otaka H, Kira M, Yano K, Ito S, Mitekura H, Kawata T and Matsui F 2004 Multi-colored dye-sensitized solar cells *J. Photochem. Photobiol. A Chem.* **164** 67–73

[23] Ren Y, Sun D, Cao Y, Tsao H N, Yuan Y, Zakeeruddin S M, Wang P and Grätzel M 2018 A Stable Blue Photosensitizer for Color Palette of Dye-Sensitized Solar Cells Reaching 12.6% Efficiency *J. Am. Chem. Soc.* **140** 2405–8

[24] Yum J-H, Holcombe T W, Kim Y, Rakstys K, Moehl T, Teuscher J, Delcamp J H, Nazeeruddin M K and Grätzel M 2013 Blue-Coloured Highly Efficient Dye-Sensitized Solar Cells by Implementing the Diketopyrrolopyrrole Chromophore *Sci. Rep.* **3** 2446

[25] Shen Z, Xu B, Liu P, Hu Y, Yu Y, Ding H, Kloo L, Hua J, Sun L and Tian H 2017 High performance solid-state dye-sensitized solar cells based on organic blue-colored dyes *J. Mater. Chem. A* **5** 1242–7

[26] Tagliaferro R, Colonna D, Brown T M, Reale A and Di Carlo A 2013 Interplay between transparency and efficiency in dye sensitized solar cells *Opt. Express* **21** 3235–42

[27] http://actu.epfl.ch/news/epfl-s-campus-has-the-world-s-first-solar-window/ (accesed on 01/10/2018)

[28] De Rossi F, Pontecorvo T and Brown T M 2015 Characterization of photovoltaic devices for indoor light harvesting and customization of flexible dye solar cells to deliver superior efficiency under artificial lighting *Appl. Energy* **156** 413–22

[29] Yoon S, Tak S, Kim J, Jun Y, Kang K and Park J 2011 Application of transparent dye-sensitized solar cells to building integrated photovoltaic systems *Build. Environ.* **46** 1899–904

[30] Reale A, Cinà L, Malatesta A, De Marco R, Brown T M and Di Carlo A 2014 Estimation of Energy Production of Dye-Sensitized Solar Cell Modules for Building-Integrated Photovoltaic Applications *Energy Technol.* **2** 531–41

[31] Ye M, Wen X, Wang M, Iocozzia J, Zhang N, Lin C and Lin Z 2015 Recent advances in dye-sensitized solar cells: from photoanodes, sensitizers and electrolytes to counter electrodes *Mater. Today* **18** 155–62

[32] Mor G K, Shankar K, Paulose M, Varghese O K and Grimes C A 2006 Use of Highly-Ordered TiO2 Nanotube Arrays in Dye-Sensitized Solar Cells *Nano Lett.* **6** 215–8

[33] Wang Z-S, Kawauchi H, Kashima T and Arakawa H 2004 Significant influence of TiO2 photoelectrode morphology on the energy conversion efficiency of N719 dye-sensitized solar cell *Coord. Chem. Rev.* **248** 1381–9

[34] Katoh R, Furube A, Yoshihara T, Hara K, Fujihashi G, Takano S, Murata S, Arakawa H and Tachiya M 2004 Efficiencies of Electron Injection from Excited N3 Dye into Nanocrystalline Semiconductor (ZrO2, TiO2, ZnO, Nb2O5, SnO2, In2O3) Films *J. Phys. Chem. B* **108** 4818–22

[35] Vittal R and Ho K-C 2017 Zinc oxide based dye-sensitized solar cells: A review *Renew. Sustain. Energy Rev.* **70** 920–35

[36] Lin C-Y, Lai Y-H, Chen H-W, Chen J-G, Kung C-W, Vittal R and Ho K-C 2011 Highly efficient dye-sensitized solar cell with a ZnO nanosheet-based photoanode *Energy Environ. Sci.* **4** 3448–55

[37] Lee J-H, Park N-G and Shin Y-J 2011 Nano-grain SnO2 electrodes for high conversion efficiency SnO2–DSSC *Sol. Energy Mater. Sol. Cells* **95** 179–83

[38] Dong Z, Ren H, Hessel C M, Wang J, Yu R, Jin Q, Yang M, Hu Z, Chen Y, Tang Z, Zhao H and Wang D 2013 Quintuple-Shelled SnO2 Hollow Microspheres with Superior Light Scattering for High-Performance Dye-Sensitized Solar Cells *Adv. Mater.* **26** 905–9

[39] Mahalingam S and Abdullah H 2016 Electron transport study of indium oxide as photoanode in DSSCs: A review *Renew. Sustain. Energy Rev.* **63** 245–55

[40] Lenzmann F, Krueger J, Burnside S, Brooks K, Grätzel M, Gal D, Rühle S and Cahen D 2001 Surface Photovoltage Spectroscopy of Dye-Sensitized Solar Cells with TiO2, Nb2O5, and SrTiO3 Nanocrystalline Photoanodes: Indication for Electron Injection from Higher Excited Dye States *J. Phys. Chem. B* **105** 6347–52

[41] Chen S G, Chappel S, Diamant Y and Zaban A 2001 Preparation of Nb2O5 Coated TiO2 Nanoporous Electrodes and Their Application in Dye-Sensitized Solar Cells *Chem. Mater.* **13** 4629–34

[42] Lee C-P, Lin R Y-Y, Lin L-Y, Li C-T, Chu T-C, Sun S-S, Lin J T and Ho K-C 2015 Recent progress in organic sensitizers for dye-sensitized solar cells *RSC Adv.* **5** 23810–25

[43] Kay A and Graetzel M 1993 Artificial photosynthesis. 1. Photosensitization of titania solar cells with chlorophyll derivatives and related natural porphyrins *J. Phys. Chem.* **97** 6272–7

[44] Calogero G, Yum J-H, Sinopoli A, Di Marco G, Grätzel M and Nazeeruddin M K 2012 Anthocyanins and betalains as light-harvesting pigments for dye-sensitized solar cells *Sol. Energy* **86** 1563–75

[45] Calogero G and Marco G Di 2008 Red Sicilian orange and purple eggplant fruits as natural sensitizers for dye-sensitized solar cells *Sol. Energy Mater. Sol. Cells* **92** 1341–6

[46] Mathew S, Yella A, Gao P, Humphry-Baker R, Curchod B F E, Ashari-Astani N, Tavernelli I, Rothlisberger U, Nazeeruddin M K and Grätzel M 2014 Dye-sensitized solar cells with 13% efficiency achieved through the molecular engineering of porphyrin sensitizers *Nat. Chem.* **6** 242

[47] Nazeeruddin M K, Péchy P, Renouard T, Zakeeruddin S M, Humphry-Baker R, Comte P, Liska P, Cevey L, Costa E, Shklover V, Spiccia L, Deacon G B, Bignozzi C A and Grätzel M 2001 Engineering of Efficient Panchromatic Sensitizers for Nanocrystalline TiO2-Based Solar Cells *J. Am. Chem. Soc.* **123** 1613–24

[48] Qin Y and Peng Q 2012 Ruthenium sensitizers and their applications in dye-sensitized solar cells *Int. J. Photoenergy* **2012** 291579

[49] Nazeeruddin M K, De Angelis F, Fantacci S, Selloni A, Viscardi G, Liska P, Ito S, Takeru B and Grätzel M 2005 Combined Experimental and DFT-TDDFT Computational Study of Photoelectrochemical Cell Ruthenium Sensitizers *J. Am. Chem. Soc.* **127** 16835–47

[50] Wu J, Lan Z, Lin J, Huang M, Huang Y, Fan L and Luo G 2015 Electrolytes in Dye-Sensitized Solar Cells *Chem. Rev.* **115** 2136–73

[51] Yu Z, Vlachopoulos N, Gorlov M and Kloo L 2011 Liquid electrolytes for dye-sensitized solar cells *Dalt. Trans.* **40** 10289–303

[52] Xia J, Masaki N, Lira-Cantu M, Kim Y, Jiang K and Yanagida S 2008 Influence of Doped Anions on Poly(3,4-ethylenedioxythiophene) as Hole Conductors for Iodine-Free Solid-State Dye-Sensitized Solar Cells *J. Am. Chem. Soc.* **130** 1258–63

[53] Nusbaumer H, Moser J-E, Zakeeruddin S M, Nazeeruddin M K and Grätzel M 2001 CoII(dbbip)22+

Complex Rivals Tri-iodide/Iodide Redox Mediator in Dye-Sensitized Photovoltaic Cells *J. Phys. Chem. B* **105** 10461–4

[54] Hamann T W, Brunschwig B S and Lewis N S 2006 Comparison of the Self-Exchange and Interfacial Charge-Transfer Rate Constants for Methyl- versus tert-Butyl-Substituted Os(III) Polypyridyl Complexes *J. Phys. Chem. B* **110** 25514–20

[55] Wu J, Lan Z, Lin J, Huang M, Huang Y, Fan L, Luo G, Lin Y, Xie Y and Wei Y 2017 Counter electrodes in dye-sensitized solar cells *Chem. Soc. Rev.* **46** 5975–6023

[56] Thomas S, Deepak T G, Anjusree G S, Arun T A, Nair S V and Nair A S 2014 A review on counter electrode materials in dye-sensitized solar cells *J. Mater. Chem. A* **2** 4474–90

[57] Calogero G, Calandra P, Irrera A, Sinopoli A, Citro I and Di Marco. G 2011 A new type of transparent and low cost counter-electrode based on platinum nanoparticles for dye-sensitized solar cells *Energy Environ. Sci.* **4** 1838–44

[58] Zhang D W, Li X D, Li H B, Chen S, Sun Z, Yin X J and Huang S M 2011 Graphene-based counter electrode for dye-sensitized solar cells *Carbon N. Y.* **49** 5382–8

[59] Ramasamy E, Lee W J, Lee D Y and Song J S 2008 Spray coated multi-wall carbon nanotube counter electrode for tri-iodide (I3-) reduction in dye-sensitized solar cells *Electrochem. commun.* **10** 1087–9

[60] Lee W J, Ramasamy E, Lee D Y and Song J S 2008 Grid type dye-sensitized solar cell module with carbon counter electrode *J. Photochem. Photobiol. A Chem.* **194** 27–30

[61] Huang Z, Liu X, Li K, Li D, Luo Y, Li H, Song W, Chen L and Meng Q 2007 Application of carbon materials as counter electrodes of dye-sensitized solar cells *Electrochem. commun.* **9** 596–8

[62] Lee W J, Ramasamy E, Lee D Y and Song J S 2008 Performance variation of carbon counter electrode based dye-sensitized solar cell *Sol. Energy Mater. Sol. Cells* **92** 814–8

[63] Calogero G, Calandra P, Sinopoli A and Gucciardi P G 2010 Metal nanoparticles and carbon-based nanostructures as advanced materials for cathode application in dye-sensitized solar cells *Int. J. Photoenergy* **2010** 109495

[64] Monreal-Bernal A, Vilatela J J and Costa R D 2019 CNT fibres as dual counter-electrode/current-collector in highly efficient and stable dye-sensitized solar cells *Carbon N. Y.* **141** 488–96

[65] Li S, Luo Y, Lv W, Yu W, Wu S, Hou P, Yang Q, Meng Q, Liu C and Cheng H-M 2011 Vertically Aligned Carbon Nanotubes Grown on Graphene Paper as Electrodes in Lithium-Ion Batteries and Dye-Sensitized Solar Cells *Adv. Energy Mater.* **1** 486–90

[66] Pereira A I, Martins J, Tavares C J, Andrade L and Mendes A 2017 Development of stable current collectors for large area dye-sensitized solar cells *Appl. Surf. Sci.* **423** 549–56

[67] Chua J, Mathews N, Jennings J R, Yang G, Wang Q and Mhaisalkar S G 2011 Patterned 3-dimensional metal grid electrodes as alternative electron collectors in dye-sensitized solar cells *Phys. Chem. Chem. Phys.* **13** 19314–7

[68] Koo B-R, Oh D-H, Riu D-H and Ahn H-J 2017 Improvement of Transparent Conducting Performance on Oxygen-Activated Fluorine-Doped Tin Oxide Electrodes Formed by Horizontal Ultrasonic Spray Pyrolysis Deposition *ACS Appl. Mater. Interfaces* **9** 44584–92

[69] Rakhshani A E, Makdisi Y and Ramazaniyan H A 1998 Electronic and optical properties of fluorine-doped tin oxide films *J. Appl. Phys.* **83** 1049–57

[70] Chowdhury F I, Blaine T and Gougam A B 2013 Optical Transmission Enhancement of Fluorine Doped Tin Oxide (FTO) on Glass for Thin Film Photovoltaic Applications *Energy Procedia* **42** 660–9

[71] Yella A, Lee H-W, Tsao H N, Yi C, Chandiran A K, Nazeeruddin M K, Diau E W-G, Yeh C-Y, Zakeeruddin S M and Grätzel M 2011 Porphyrin-Sensitized Solar Cells with Cobalt (II/III)–Based Redox Electrolyte Exceed 12 Percent Efficiency *Science* **334** 629-634

[72] Yella A, Mathew S, Aghazada S, Comte P, Grätzel M and Nazeeruddin M K 2017 Dye-sensitized solar cells using cobalt electrolytes: the influence of porosity and pore size to achieve high-efficiency *J. Mater. Chem. C* **5** 2833–43

[73] Bella F, Galliano S, Gerbaldi C and Viscardi G 2016 Cobalt-Based Electrolytes for Dye-Sensitized Solar Cells: Recent Advances towards Stable Devices *Energies* **9** 384

[74] Ghamouss F, Pitson R, Odobel F, Boujtita M, Caramori S and Bignozzi C A 2010 Characterization of screen printed carbon counter electrodes for Co(II)/(III) mediated photoelectrochemical cells *Electrochim. Acta* **55** 6517–6522

[75] Feldt S M, Gibson E A, Gabrielsson E, Sun L, Boschloo G and Hagfeldt A 2010 Design of Organic Dyes and Cobalt Polypyridine Redox Mediators for High-Efficiency Dye-Sensitized Solar Cells *J. Am. Chem. Soc.* **132** 16714–24

[76] Feldt S M, Wang G, Boschloo G and Hagfeldt A 2011 Effects of Driving Forces for Recombination and Regeneration on the Photovoltaic Performance of Dye-Sensitized Solar Cells using Cobalt Polypyridine Redox Couples *J. Phys. Chem. C* **115** 21500–7

[77] Daeneke T, Kwon T-H, Holmes A B, Duffy N W, Bach U and Spiccia L 2011 High-efficiency dye-sensitized solar cells with ferrocene-based electrolytes *Nat. Chem.* **3** 211

[78] Boschloo G and Hagfeldt A 2009 Characteristics of the Iodide/Triiodide Redox Mediator in Dye-Sensitized Solar Cells *Acc. Chem. Res.* **42** 1819–26

[79] Bessho T, Yoneda E, Yum J-H, Guglielmi M, Tavernelli I, Imai H, Rothlisberger U, Nazeeruddin M K and Grätzel M 2009 New Paradigm in Molecular Engineering of Sensitizers for Solar Cell Applications *J. Am. Chem. Soc.* **131** 5930–4

[80] Hamann T W and Ondersma J W 2011 Dye-sensitized solar cell redox shuttles *Energy Environ. Sci.* **4** 370–81

[81] Zardetto V, Brown T M, Reale A and Di Carlo A 2011 Substrates for flexible electronics: A practical investigation on the electrical, film flexibility, optical, temperature, and solvent resistance properties *J.*

[82] Minami T 2005 Transparent conducting oxide semiconductors for transparent electrodes *Semicond. Sci. Technol.* **20** S35

[83] Tseng S-F, Hsiao W-T, Chiang D, Huang K-C and Chou C-P 2011 Mechanical and optoelectric properties of post-annealed fluorine-doped tin oxide films by ultraviolet laser irradiation *Appl. Surf. Sci.* **257** 7204–9

[84] Ellmer K 2012 Past achievements and future challenges in the development of optically transparent electrodes *Nat. Photonics* **6** 809

[85] Kumar A and Zhou C 2010 The Race To Replace Tin-Doped Indium Oxide: Which Material Will Win? *ACS Nano* **4** 11–4

[86] Huang X, Yu Z, Huang S, Zhang Q, Li D, Luo Y and Meng Q 2010 Preparation of fluorine-doped tin oxide (SnO2:F) film on polyethylene terephthalate (PET) substrate *Mater. Lett.* **64** 1701–3

[87] http://www.platinum.matthey.com/prices/price-tables (accessed on 01/09/2018)

[88] Koo B-K, Lee D-Y, Kim H-J, Lee W-J, Song J-S and Kim H-J 2006 Seasoning effect of dye-sensitized solar cells with different counter electrodes *J. Electroceramics* **17** 79–82

[89] Syrrokostas G, Siokou A, Leftheriotis G and Yianoulis P 2012 Degradation mechanisms of Pt counter electrodes for dye sensitized solar cells *Sol. Energy Mater. Sol. Cells* **103** 119–27

[90] Imoto K, Takahashi K, Yamaguchi T, Komura T, Nakamura J and Murata K 2003 High-performance carbon counter electrode for dye-sensitized solar cells *Sol. Energy Mater. Sol. Cells* **79** 459–69

[91] Calogero G, Bonaccorso F, Maragò O M, Gucciardi P G and Di Marco G 2010 Single wall carbon nanotubes deposited on stainless steel sheet substrates as novel counter electrodes for ruthenium polypyridine based dye sensitized solar cells *Dalt. Trans.* **39** 2903–9

[92] Trancik J E, Barton S C and Hone J 2008 Transparent and Catalytic Carbon Nanotube Films *Nano Lett.* **8** 982–7

[93] Veerappan G, Bojan K and Rhee S-W 2011 Sub-micrometer-sized Graphite As a Conducting and Catalytic Counter Electrode for Dye-sensitized Solar Cells *ACS Appl. Mater. Interfaces* **3** 857–62

[94] Chou H, Lien C, Wu D, Hsu H and Huang M 2013 The different film thicknesses of graphite/carbon black on the counter electrodes by spray coating method for dye-sensitized solar cells *2013 13th IEEE International Conference on Nanotechnology (IEEE-NANO 2013)* pp 1151–4

[95] Kay A and Grätzel M 1996 Low cost photovoltaic modules based on dye sensitized nanocrystalline titanium dioxide and carbon powder *Sol. Energy Mater. Sol. Cells* **44** 99–117

[96] Roy-Mayhew J D, Bozym D J, Punckt C and Aksay I A 2010 Functionalized Graphene as a Catalytic Counter Electrode in Dye-Sensitized Solar Cells *ACS Nano* **4** 6203–11

[97] Ju M J, Kim J C, Choi H-J, Choi I T, Kim S G, Lim K, Ko J, Lee J-J, Jeon I-Y, Baek J-B and Kim H K 2013 N-Doped Graphene Nanoplatelets as Superior Metal-Free Counter Electrodes for Organic Dye-Sensitized Solar Cells *ACS Nano* **7** 5243–50

[98] Kavan L, Yum J-H and Graetzel M 2012 Optically Transparent Cathode for Co(III/II) Mediated Dye-Sensitized Solar Cells Based on Graphene Oxide *ACS Appl. Mater. Interfaces* **4** 6999–7006

[99] Choi H, Kim H, Hwang S, Choi W and Jeon M 2011 Dye-sensitized solar cells using graphene-based carbon nano composite as counter electrode *Sol. Energy Mater. Sol. Cells* **95** 323–5

[100] Zakhidov J V and A J M and D L and D O and G W and R B and A 2012 Carbon nanotube/graphene nanocomposite as efficient counter electrodes in dye-sensitized solar cells *Nanotechnology* **23** 85201

[101] Kavan L, Yum J-H and Grätzel M 2011 Graphene Nanoplatelets Outperforming Platinum as the Electrocatalyst in Co-Bipyridine-Mediated Dye-Sensitized Solar Cells *Nano Lett.* **11** 5501–6

[102] Roy S, Bajpai R, Jena A K, Kumar P, kulshrestha N and Misra D S 2012 Plasma modified flexible bucky paper as an efficient counter electrode in dye sensitized solar cells *Energy Environ. Sci.* **5** 7001–6

[103] Anothumakkool B, Agrawal I, Bhange S N, Soni R, Game O, Ogale S B and Kurungot S 2016 Pt- and TCO-Free Flexible Cathode for DSSC from Highly Conducting and Flexible PEDOT Paper Prepared via in Situ Interfacial Polymerization *ACS Appl. Mater. Interfaces* **8** 553–62

[104] Geim A K and Novoselov K S 2007 The rise of graphene *Nat. Mater.* **6** 183–91

[105] Zhu Y, Murali S, Cai W, Li X, Suk J W, Potts J R and Ruoff R S 2010 Graphene and Graphene Oxide: Synthesis, Properties, and Applications *Adv. Mater.* **22** 3906–24

[106] Chen J H, Jang C, Xiao S, Ishigami M and Fuhrer M S 2008 Intrinsic and extrinsic performance limits of graphene devices on SiO2 *Nat. Nanotechnol.* **3** 206

[107] Bolotin K I, Sikes K J, Jiang Z, Klima M, Fudenberg G, Hone J, Kim P and Stormer H L 2008 Ultrahigh electron mobility in suspended graphene *Solid State Commun.* **146** 351–5

[108] Wassei J K and Kaner R B 2010 Graphene, a promising transparent conductor *Mater. Today* **13** 52–9

[109] De S and Coleman J N 2010 Are There Fundamental Limitations on the Sheet Resistance and Transmittance of Thin Graphene Films? *ACS Nano* **4** 2713–20

[110] Balandin A A 2011 Thermal properties of graphene and nanostructured carbon materials *Nat. Mater.* **10** 569

[111] Balandin D L N and A A 2012 Two-dimensional phonon transport in graphene *J. Phys. Condens. Matter* **24** 233203

[112] Papageorgiou D G, Kinloch I A and Young R J 2017 Mechanical properties of graphene and graphene-based nanocomposites *Prog. Mater. Sci.* **90** 75–127

[113] Lee C, Wei X, Kysar J W and Hone J 2008 Measurement of the Elastic Properties and Intrinsic Strength of Monolayer Graphene *Science* **321** 385

[114] Najafi L, Bellani S, Oropesa-Nuñez R, Ansaldo A, Prato M, Del Rio Castillo A and Bonaccordo F 2018 Engineered MoSe2-Based Heterostructures for Efficient Electrochemical Hydrogen Evolution Reaction *Adv. Energy Mater.* **8** 1703212

[115]  Huang C, Li C and Shi G 2012 Graphene based catalysts *Energy Environ. Sci.* **5** 8848–68
[116]  Machado B F and Serp P 2012 Graphene-based materials for catalysis *Catal. Sci. Technol.* **2** 54–75
[117]  Kong X-K, Chen C-L and Chen Q-W 2014 Doped graphene for metal-free catalysis *Chem. Soc. Rev.* **43** 2841–57
[118]  Navalon S, Dhakshinamoorthy A, Alvaro M and Garcia H 2014 Carbocatalysis by Graphene-Based Materials *Chem. Rev.* **114** 6179–212
[119]  Huang X, Zeng Z, Fan Z, Liu J and Zhang H 2012 Graphene-based electrodes *Adv. Mater.* **24** 5979–6004
[120]  Xu Y and Liu J 2016 Graphene as Transparent Electrodes: Fabrication and New Emerging Applications *Small* **12** 1400–19
[121]  Kim K S, Zhao Y, Jang H, Lee S Y, Kim J M, Kim K S, Ahn J-H, Kim P, Choi J-Y and Hong B H 2009 Large-scale pattern growth of graphene films for stretchable transparent electrodes *Nature* **457** 706–10
[122]  Bonaccorso F, Sun Z, Hasan T and Ferrari A C 2010 Graphene photonics and optoelectronics *Nat. Photonics* **4** 611–22
[123]  Bonaccorso F, Colombo L, Yu G, Stoller M, Tozzini V, Ferrari a C, Ruoff R S and Pellegrini V 2015 2D materials. Graphene, related two-dimensional crystals, and hybrid systems for energy conversion and storage *Science* **347** 1246501
[124]  Wang X, Zhi L and Müllen K 2008 Transparent, Conductive Graphene Electrodes for Dye-Sensitized Solar Cells *Nano Lett.* **8** 323–7
[125]  Wu J, Becerril H A, Bao Z, Liu Z, Chen Y and Peumans P 2008 Organic solar cells with solution-processed graphene transparent electrodes *Appl. Phys. Lett.* **92** 263302
[126]  Yoon J, Sung H, Lee G, Cho W, Ahn N, Jung H S and Choi M 2017 Superflexible, high-efficiency perovskite solar cells utilizing graphene electrodes: towards future foldable power sources *Energy Environ. Sci.* **10** 337–45
[127]  Capasso A, Salamandra L, Faggio G, Dikonimos T, Buonocore F, Morandi V, Ortolani L and Lisi N 2016 Chemical Vapor Deposited Graphene-Based Derivative As High-Performance Hole Transport Material for Organic Photovoltaics *ACS Appl. Mater. Interfaces* **8** 23844–53
[128]  Lancellotti L, Bobeico E, Capasso A, Lago E, Delli Veneri P, Leoni E, Buonocore F and Lisi N 2016 Combined effect of double antireflection coating and reversible molecular doping on performance of few-layer graphene/n-silicon Schottky barrier solar cells *Sol. Energy* **127** 198–205
[129]  Li N, Oida S, Tulevski G S, Han S-J, Hannon J B, Sadana D K and Chen T-C 2013 Efficient and bright organic light-emitting diodes on single-layer graphene electrodes *Nat. Commun.* **4** 2294
[130]  Han T-H, Lee Y, Choi M-R, Woo S-H, Bae S-H, Hong B H, Ahn J-H and Lee T-W 2012 Extremely efficient flexible organic light-emitting diodes with modified graphene anode *Nat. Photonics* **6** 105
[131]  Wu J, Agrawal M, Becerril H A, Bao Z, Liu Z, Chen Y and Peumans P 2010 Organic Light-Emitting Diodes on Solution-Processed Graphene Transparent Electrodes *ACS Nano* **4** 43–8
[132]  Bae S, Kim H, Lee Y, Xu X, Park J-S, Zheng Y, Balakrishnan J, Lei T, Ri Kim H, Song Y II, Kim Y-J, Kim K S, Özyilmaz B, Ahn J-H, Hong B H and Iijima S 2010 Roll-to-roll production of 30-inch graphene films for transparent electrodes *Nat. Nanotechnol.* **5** 574
[133]  Khan U, Kim T-H, Ryu H, Seung W and Kim S-W 2016 Graphene Tribotronics for Electronic Skin and Touch Screen Applications *Adv. Mater.* **29** 1603544
[134]  Ahn J-H and Hong B H 2014 Graphene for displays that bend *Nat. Nanotechnol.* **9** 737
[135]  Liu N, Chortos A, Lei T, Jin L, Kim T R, Bae W-G, Zhu C, Wang S, Pfattner R, Chen X, Sinclair R and Bao Z 2017 Ultratransparent and stretchable graphene electrodes *Sci. Adv.* **3**
[136]  Novoselov K S, Geim A K, Morozov S V, Jiang D, Zhang Y, Dubonos S V, Grigorieva I V and Firsov A A 2004 Electric Field Effect in Atomically Thin Carbon Films *Science* **306** 666–9
[137]  Koppens F H L, Mueller T, Avouris P, Ferrari A C, Vitiello M S and Polini M 2014 Photodetectors based on graphene, other two-dimensional materials and hybrid systems *Nat. Nanotechnol.* **9** 780–93
[138]  Pumera M 2011 Graphene in biosensing *Mater. Today* **14** 308–15
[139]  Shao Y, Wang J, Wu H, Liu J, Aksay I A and Lin Y 2010 Graphene Based Electrochemical Sensors and Biosensors: A Review *Electroanalysis* **22** 1027–36
[140]  Pumera M, Ambrosi A, Bonanni A, Chng E L K and Poh H L 2010 Graphene for electrochemical sensing and biosensing *TrAC Trends Anal. Chem.* **29** 954–65
[141]  Xu C, Xu B, Gu Y, Xiong Z, Sun J and Zhao X S 2013 Graphene-based electrodes for electrochemical energy storage *Energy Environ. Sci.* **6** 1388–414
[142]  Liu C, Yu Z, Neff D, Zhamu A and Jang B Z 2010 Graphene-Based Supercapacitor with an Ultrahigh Energy Density *Nano Lett.* **10** 4863–8
[143]  Wang Y, Shi Z, Huang Y, Ma Y, Wang C, Chen M and Chen Y 2009 Supercapacitor Devices Based on Graphene Materials *J. Phys. Chem. C* **113** 13103–7
[144]  Reddy A L M, Srivastava A, Gowda S R, Gullapalli H, Dubey M and Ajayan P M 2010 Synthesis Of Nitrogen-Doped Graphene Films For Lithium Battery Application *ACS Nano* **4** 6337–42
[145]  Ansaldo A, Bondavalli P, Bellani S, Del Rio Castillo A E, Prato M, Pellegrini V, Pognon G and Bonaccorso F 2017 High-power graphene–Carbon nanotube hybrid supercapacitors *ChemNanoMat* **3** 436
[146]  Liu J 2014 Charging graphene for energy *Nat. Nanotechnol.* **9** 739
[147]  Bonaccorso F, Lombardo A, Hasan T, Sun Z, Colombo L and Ferrari A C 2012 Production and processing of graphene and 2d crystals *Mater. Today* **15** 564–89
[148]  Nicolosi V, Chhowalla M, Kanatzidis M G, Strano M S and Coleman J N 2013 Liquid Exfoliation of Layered Materials *Science* **340** 1226419
[149]  Ren W and Cheng H-M 2014 The global growth of graphene *Nat. Nanotechnol.* **9** 726
[150]  Tung V C, Allen M J, Yang Y and Kaner R B 2008 High-throughput solution processing of large-scale graphene *Nat. Nanotechnol.* **4** 25
[151]  Bonaccorso F, Bortolotta A, Coleman J N and Backes C 2016 2D-Crystal-Based Functional Inks

*Adv. Mater.* **28** 6136–66
[152]  Youn D H, Jang J-W, Kim J Y, Jang J S, Choi S H and Lee J S 2014 Fabrication of graphene-based electrode in less than a minute through hybrid microwave annealing *Sci. Rep.* **4** 5492
[153]  Li D, Müller M B, Gilje S, Kaner R B and Wallace G G 2008 Processable aqueous dispersions of graphene nanosheets *Nat. Nanotechnol.* **3** 101
[154]  Ciriminna R, Zhang N, Yang M-Q, Meneguzzo F, Xu Y-J and Pagliaro M 2015 Commercialization of graphene-based technologies: a critical insight *Chem. Commun.* **51** 7090–5
[155]  Muñoz R and Gómez-Aleixandre C 2013 Review of CVD Synthesis of Graphene *Chem. Vap. Depos.* **19** 297–322
[156]  Mattevi C, Kim H and Chhowalla M 2011 A review of chemical vapour deposition of graphene on copper *J. Mater. Chem.* **21** 3324–34
[157]  Faggio G, Capasso A, Messina G, Santangelo S, Dikonimos T, Gagliardi S, Giorgi R, Morandi V, Ortolani L and Lisi N 2013 High-Temperature Growth of Graphene Films on Copper Foils by Ethanol Chemical Vapor Deposition *J. Phys. Chem. C* **117** 21569–76
[158]  Lisi N, Dikonimos T, Buonocore F, Pittori M, Mazzaro R, Rizzoli R, Marras S and Capasso A 2017 Contamination-free graphene by chemical vapor deposition in quartz furnaces *Sci. Rep.* **7** 9927
[159]  Kang J, Shin D, Bae S and Hong B H 2012 Graphene transfer: key for applications *Nanoscale* **4** 5527–37
[160]  Chandrashekar B N, Deng B, Smitha A S, Chen Y, Tan C, Zhang H, Peng H and Liu Z 2015 Roll-to-Roll Green Transfer of CVD Graphene onto Plastic for a Transparent and Flexible Triboelectric Nanogenerator *Adv. Mater.* **27** 5210–6
[161]  Deokar G, Avila J, Razado-Colambo I, Codron J-L, Boyaval C, Galopin E, Asensio M-C and Vignaud D 2015 Towards high quality CVD graphene growth and transfer *Carbon N. Y.* **89** 82–92
[162]  Reina A, Jia X, Ho J, Nezich D, Son H, Bulovic V, Dresselhaus M S and Kong J 2009 Large Area, Few-Layer Graphene Films on Arbitrary Substrates by Chemical Vapor Deposition *Nano Lett.* **9** 30–5
[163]  Li X, Cai W, An J, Kim S, Nah J, Yang D, Piner R, Velamakanni A, Jung I, Tutuc E, Banerjee S K, Colombo L and Ruoff R S 2009 Large-Area Synthesis of High-Quality and Uniform Graphene Films on Copper Foils *Science* **324** 1312
[164]  Capasso A, De Francesco M, Leoni E, Dikonimos T, Buonocore F, Lancellotti L, Bobeico E, Sarto M S, Tamburrano A, De Bellis G and Lisi N 2014 Cyclododecane as support material for clean and facile transfer of large-area few-layer graphene *Appl. Phys. Lett.* **105** 113101
[165]  Caldwell J D, Anderson T J, Culbertson J C, Jernigan G G, Hobart K D, Kub F J, Tadjer M J, Tedesco J L, Hite J K, Mastro M A, Myers-Ward R L, Eddy C R, Campbell P M and Gaskill D K 2010 Technique for the Dry Transfer of Epitaxial Graphene onto Arbitrary Substrates *ACS Nano* **4** 1108–14
[166]  Grande M, Bianco G V, Vincenti M A, de Ceglia D, Capezzuto P, Scalora M, D'Orazio A and Bruno G 2015 Optically Transparent Microwave Polarizer Based On Quasi-Metallic Graphene *Sci. Rep.* **5** 17083
[167]  La Notte L, Villari E, Palma A L, Sacchetti A, Michela Giangregorio M, Bruno G, Di Carlo A, Bianco G V and Reale A 2017 Laser-patterned functionalized CVD-graphene as highly transparent conductive electrodes for polymer solar cells *Nanoscale* **9** 62–9
[168]  Gordon R G 2000 Criteria for Choosing Transparent Conductors *MRS Bull.* **25** 52–7
[169]  Minami T 2008 Present status of transparent conducting oxide thin-film development for Indium-Tin-Oxide (ITO) substitutes *Thin Solid Films* **516** 5822–8
[170]  Bellani S, Najafi L, Tullii G, Ansaldo A, Oropesa-Nuñez R, Prato M, Colombo M, Antognazza M R and Bonaccorso F 2017 ITO nanoparticles break optical transparency/high-areal capacitance trade-off for advanced aqueous supercapacitors *J. Mater. Chem. A* **5** 25177
[171]  Mryasov O N and Freeman A J 2001 Electronic band structure of indium tin oxide and criteria for transparent conducting behavior *Phys. Rev. B* **64** 233111
[172]  Andersson A, Johansson N, Bröms P, Yu N, Lupo D and Salaneck W R 1999 Fluorine Tin Oxide as an Alternative to Indium Tin Oxide in Polymer LEDs *Adv. Mater.* **10** 859–63
[173]  Bonaccorso F, Bartolotta A, Coleman J N and Backes C 2016 2D-Crystal-Based Functional Inks *Adv. Mater.* 6136–66
[174]  Hernandez Y, Nicolosi V, Lotya M, Blighe F M, Sun Z, De S, McGovern I T, Holland B, Byrne M, Gun'Ko Y K, Boland J J, Niraj P, Duesberg G, Krishnamurthy S, Goodhue R, Hutchison J, Scardaci V, Ferrari A C and Coleman J N 2008 High-yield production of graphene by liquid-phase exfoliation of graphite *Nat. Nanotechnol.* **3** 563
[175]  Del Rio Castillo A E E, Pellegrini V, Ansaldo A, Ricciardella F, Sun H, Marasco L, Buha J, Dang Z, Gagliani L, Lago E, Curreli N, Gentiluomo S, Palazon F, Prato M, Oropesa-Nunez R, Toth P, Mantero E, Crugliano M, Gamucci A, Tomadin A, Polini M and Bonaccorso F 2018 High-yield production of 2D crystals by wet-jet milling *Mater. Horizons* **5** 890
[176]  Ciesielski A and Samori P 2014 Graphene via sonication assisted liquid-phase exfoliation *Chem. Soc. Rev.* **43** 381–98
[177]  Khan U, O'Neill A, Lotya M, De S and Coleman J N 2010 High-Concentration Solvent Exfoliation of Graphene *Small* **6** 864–71
[178]  Shih C-J, Vijayaraghavan A, Krishnan R, Sharma R, Han J-H, Ham M-H, Jin Z, Lin S, Paulus G L C, Reuel N F, Wang Q H, Blankschtein D and Strano M S 2011 Bi- and trilayer graphene solutions *Nat. Nanotechnol.* **6** 439
[179]  Lotya M, King P J, Khan U, De S and Coleman J N 2010 High-Concentration, Surfactant-Stabilized Graphene Dispersions *ACS Nano* **4** 3155–62
[180]  Hasan T, Torrisi F, Sun Z, Popa D, Nicolosi V, Privitera G, Bonaccorso F and Ferrari A C A C 2010 Solution-phase exfoliation of graphite for ultrafast photonics **247** 2953–7
[181]  Notley S M 2012 Highly Concentrated Aqueous Suspensions of Graphene through Ultrasonic

Exfoliation with Continuous Surfactant Addition *Langmuir* **28** 14110–3
[182]  Zhao W, Fang M, Wu F, Wu H, Wang L and Chen G 2010 Preparation of graphene by exfoliation of graphite using wet ball milling *J. Mater. Chem.* **20** 5817–9
[183]  Jeon I-Y, Shin Y-R, Sohn G-J, Choi H-J, Bae S-Y, Mahmood J, Jung S-M, Seo J-M, Kim M-J, Wook Chang D, Dai L and Baek J-B 2012 Edge-carboxylated graphene nanosheets via ball milling *Proc. Natl. Acad. Sci.* **109** 5588
[184]  Paton K R, Varrla E, Backes C, Smith R J, Khan U, O'Neill A, Boland C, Lotya M, Istrate O M, King P, Higgins T, Barwich S, May P, Puczkarski P, Ahmed I, Moebius M, Pettersson H, Long E, Coelho J, O'Brien S E, McGuire E K, Sanchez B M, Duesberg G S, McEvoy N, Pennycook T J, Downing C, Crossley A, Nicolosi V and Coleman J N 2014 Scalable production of large quantities of defect-free few-layer graphene by shear exfoliation in liquids *Nat. Mater.* **13** 624
[185]  Varrla E, Paton K R, Backes C, Harvey A, Smith R J, McCauley J and Coleman J N 2014 Turbulence-assisted shear exfoliation of graphene using household detergent and a kitchen blender *Nanoscale* **6** 11810–9
[186]  Tran T S, Park S J, Yoo S S, Lee T-R and Kim T 2016 High shear-induced exfoliation of graphite into high quality graphene by Taylor–Couette flow *RSC Adv.* **6** 12003–8
[187]  Blomquist N, Engström A-C, Hummelgård M, Andres B, Forsberg S and Olin H 2016 Large-Scale Production of Nanographite by Tube-Shear Exfoliation in Water *PLoS One* **11** e0154686
[188]  Bellani S, Wang F, Longoni G, Najafi L, Oropesa-Nuñez R, Del Rio Castillo A E, Prato M, Zhuang X, Pellegrini V, Feng X and Bonaccorso F 2018 WS2–Graphite Dual-Ion Batteries *Nano Lett.* **18** 7155
[189]  Capasso A, Del Rio Castillo A E, Sun H, Ansaldo A, Pellegrini V and Bonaccorso F 2015 Ink-jet printing of graphene for flexible electronics: An environmentally-friendly approach *Solid State Commun.* **224** 53–63
[190]  McManus D, Vranic S, Withers F, Sanchez-Romaguera V, Macucci M, Yang H, Sorrentino R, Parvez K, Son S K, Iannaccone G, Kostarelos K, Fiori G and Casiraghi C 2017 Water-based and biocompatible 2D crystal inks for all-inkjet-printed heterostructures *Nat. Nanotechnol.* **12** *343*
[191]  Casaluci S, Gemmi M, Pellegrini V, Di Carlo A and Bonaccorso F 2016 Graphene-based large area dye-sensitized solar cell modules *Nanoscale* **8** 5368–78
[192]  Najafi L, Bellani S, Martín-García B, Oropesa-Nuñez R, Del Rio Castillo A E, Prato M, Moreels I and Bonaccorso F 2017 Solution-Processed Hybrid Graphene Flake/2H-MoS2 Quantum Dot Heterostructures for Efficient Electrochemical Hydrogen Evolution *Chem. Mater.* **29** 5782–6
[193]  Bellani S, Najafi L, Martín-García B, Ansaldo A, Del Rio Castillo A E, Prato M, Moreels I and Bonaccorso F 2017 Graphene-Based Hole-Selective Layers for High-Efficiency, Solution-Processed, Large-Area, Flexible, Hydrogen-Evolving Organic Photocathodes *J. Phys. Chem. C* **121** 21887–903
[194]  Taheri B, Yaghoobi N N, Agresti A, Pescetelli S, Ciceroni C, Del Rio Castillo, A E, Cinà L, Bellani S, Bonaccorso F, Di Carlo A 2018 Graphene-engineered automated sprayed mesoscopic structure for perovskite device scaling-up *2D Mater.* **5** 045034
[195]  Sharma D, Jha R and Kumar S 2016 Quantum dot sensitized solar cell: Recent advances and future perspectives in photoanode *Sol. Energy Mater. Sol. Cells* **155** 294–322
[196]  Jun H K, Careem M A and Arof A K 2013 Quantum dot-sensitized solar cells—perspective and recent developments: A review of Cd chalcogenide quantum dots as sensitizers *Renew. Sustain. Energy Rev.* **22** 148–67
[197]  Ubani C A, Ibrahim M A and Teridi M A M 2017 Moving into the domain of perovskite sensitized solar cell *Renew. Sustain. Energy Rev.* **72** 907–15
[198]  Ng C H, Lim H N, Hayase S, Zainal Z, Shafie S, Lee H W and Huang N M 2018 Cesium Lead Halide Inorganic-Based Perovskite-Sensitized Solar Cell for Photo-Supercapacitor Application under High Humidity Condition *ACS Appl. Energy Mater.* **1** 692–9
[199]  Yun S, Qin Y, Uhl A R, Vlachopoulos N, Yin M, Li D, Han X and Hagfeldt A 2018 New-generation integrated devices based on dye-sensitized and perovskite solar cells *Energy Environ. Sci.* **11** 476–526
[200]  Miseikis V, Convertino D, Mishra N, Gemmi M, Mashoff T, Heun S, Haghighian N, Bisio F, Canepa M, Piazza V and Coletti C 2015 Rapid CVD growth of millimetre-sized single crystal graphene using a cold-wall reactor *2D Mater.* **2** 14006
[201]  Li X, Zhu Y, Cai W, Borysiak M, Han B, Chen D, Piner R D, Colomba L and Ruoff R S 2009 Transfer of large-area graphene films for high-performance transparent conductive electrodes *Nano Lett.* **12** 4359
[202]  Hassoun J, Bonaccorso F, Agostini M, Angelucci M, Betti M G, Cingolani R, Gemmi M, Mariani C, Panero S, Pellegrini V and Scrosati B 2014 An Advanced Lithium-Ion Battery Based on a Graphene Anode and a Lithium Iron Phosphate Cathode *Nano Lett.* **14** 4901–6
[203]  Bonaccorso F, Balis N, Stylianakis M M, Savarese M, Adamo C, Gemmi M, Pellegrini V, Stratakis E and Kymakis E 2015 Functionalized Graphene as an Electron-Cascade Acceptor for Air-Processed Organic Ternary Solar Cells *Adv. Funct. Mater.* **25** 3870
[204]  Lotya M, Hernandez Y, King P J, Smith R J, Nicolosi V, Karlsson L S, Blighe F M, De S, Wang Z, McGovern I T, Duesberg G S and Coleman J N 2009 Liquid Phase Production of Graphene by Exfoliation of Graphite in Surfactant/Water Solutions *J. Am. Chem. Soc.* **131** 3611–20
[205]  Brunauer S, Emmett P H and Teller E 1938 Adsorption of Gases in Multimolecular Layers *J. Am. Chem. Soc.* **60** 309–19
[206]  Walton K S and Snurr R Q 2007 Applicability of the BET Method for Determining Surface Areas of Microporous Metal−Organic Frameworks *J. Am. Chem. Soc.* **129** 8552–6
[207]  Fagerlund G 1973 Determination of specific surface by the BET method *Matériaux Constr.* **6** 239–45
[208]  Wang S, Minami D and Kaneko K 2015 Comparative pore structure analysis of highly porous graphene

monoliths treated at different temperatures with adsorption of N$_2$ at 77.4 K and of Ar at 87.3 K and 77.4 K *Microporous Mesoporous Mater.* **209** 72

[209] Thommes M, Kaneko K, Neimark AV, Olivier J P, Rodriguez-Reinoso F, Rouquerol J and Sing K S 2015 Physisorption of gases, with special reference to the evaluation of surface area and pore size distribution (IUPAC Technical Report) *Pure Appl. Chem.* **87** 1051

[210] Bonaccorso F, Tan P-H and Ferrari A C 2013 Multiwall Nanotubes, Multilayers, and Hybrid Nanostructures: New Frontiers for Technology and Raman Spectroscopy *ACS Nano* **7** 1838–44

[211] Han G H, Güneş F, Bae J J, Kim E S, Chae S J, Shin H J, Choi J Y, Pribat D and Lee Y H 2011 Influence of copper morphology in forming nucleation seeds for graphene growth *Nano Lett.* **11** 4144

[212] Meyer J C, Geim A K, Katsnelson M I, Novoselov K S, Booth T J and Roth S 2007 The structure of suspended graphene sheets *Nature* **446** 60

[213] Gupta A, Chen G, Joshi P, Tadigadapa S and Eklund P C 2006 Raman scattering from high-frequency phonons in supported n-graphene layer films *Nano Lett.* **6** 2667

[214] Nemes-Incze P, Osváth Z, Kamarás K and Biró L P 2008 Anomalies in thickness measurements of graphene and few layer graphite crystals by tapping mode atomic force microscopy *Carbon* **46** 1435–42

[215] Shearer C J, Slattery A D, Stapleton A J, Shapter J G and Gibson C T 2016 Accurate thickness measurement of graphene *Nanotechnology* **27** 125704

[216] Ferrari A C, Meyer J C, Scardaci V, Casiraghi C, Lazzeri M, Mauri F, Piscanec S, Jiang D, Novoselov K S, Roth S and Geim A K 2006 Raman Spectrum of Graphene and Graphene Layers *Phys. Rev. Lett.* **97** 187401

[217] Wu J-B, Lin M-L, Cong X, Liu H-N and Tan P-H 2018 Raman spectroscopy of graphene-based materials and its applications in related devices *Chem. Soc. Rev.* **47** 1822–73

[218] Ferrari A C and Basko D M 2013 Raman spectroscopy as a versatile tool for studying the properties of graphene *Nat. Nanotechnol.* **8** 235

[219] Dresselhaus M S, Dresselhaus G and Hofmann M 2008 Raman spectroscopy as a probe of graphene and carbon nanotubes *Philos. Trans. R. Soc. A Math. Phys. Eng. Sci.* **366** 231

[220] Nuvoli D, Valentini L, Alzari V, Scognamillo S, Bon S B, Piccinini M, Illescas J and Mariani A 2011 High concentration few-layer graphene sheets obtained by liquid phase exfoliation of graphite in ionic liquid *J. Mater. Chem.* **21** 3428–31

[221] Ferrari A C 2007 Raman spectroscopy of graphene and graphite: Disorder, electron-phonon coupling, doping and nonadiabatic effects *Solid State Commun.* **143** 47–57

[222] Ferrari A C and Robertson J 2000 Interpretation of Raman spectra of disordered and amorphous carbon *Phys. Rev. B* **61** 14095–107

[223] Ferrari A C and Robertson J 2001 Resonant Raman spectroscopy of disordered, amorphous, and diamondlike carbon *Phys. Rev. B* **64** 75414

[224] Yang L, Deslippe J, Park C-H, Cohen M L and Louie S G 2009 Excitonic Effects on the Optical Response of Graphene and Bilayer Graphene *Phys. Rev. Lett.* **103** 186802

[225] Belviso S, Capasso A, Santoro E, Najafi L, Lelj F, Superchi S, Casarini D, Villani C, Spirito D, Bellani S, Del Rio-Castillo A E and Bonaccorso F 2018 Thioethyl-Porphyrazine/Nanocarbon Hybrids for Photoinduced Electron Transfer *Adv. Funct. Mater.* **28** 1705418

[226] Bracamonte M V, Lacconi G I, Urreta S E and Foa Torres L E F 2014 On the Nature of Defects in Liquid-Phase Exfoliated Graphene *J. Phys. Chem. C* **118** 15455–9

[227] Coleman J N 2013 Liquid Exfoliation of Defect-Free Graphene *Acc. Chem. Res.* **46** 14–22

[228] Huang Z, Natu G, Ji Z, Hasin P and Wu Y 2011 p-Type Dye-Sensitized NiO Solar Cells: A Study by Electrochemical Impedance Spectroscopy *J. Phys. Chem. C* **115** 25109–14

[229] Chen L, Tan W, Zhang J, Zhou X, Zhang X and Lin Y 2010 Fabrication of high performance Pt counter electrodes on conductive plastic substrate for flexible dye-sensitized solar cells *Electrochim. Acta* **55** 3721–6

[230] Halme J, Vahermaa P, Miettunen K and Lund P 2010 Device Physics of Dye Solar Cells *Adv. Mater.* **22** E210–34

[231] Barsoukov E and Macdonald J R 2005 *Impedance Spectroscopy*

[232] Wang Q, Moser J-E and Grätzel M 2005 Electrochemical Impedance Spectroscopic Analysis of Dye-Sensitized Solar Cells *J. Phys. Chem. B* **109** 14945

[233] Fabregat-Santiago F, Bisquert J, Garcia-Belmonte G, Boschloo G and Hagfeldt A 2005 Influence of electrolyte in transport and recombination in dye-sensitized solar cells studied by impedance spectroscopy *Solar Energy Materials and Solar Cells* **87** 117-131

[234] Adachi M, Sakamoto M, Jiu J, Ogata Y and Isoda S 2006 Determination of parameters of electron transport in dye-sensitized solar cells using electrochemical impedance spectroscopy. *J. Phys. Chem. B* **110** 13872

[235] Hauch A and Georg A 2001 Diffusion in the electrolyte and charge-transfer reaction at the platinum electrode in dye-sensitized solar cells *Electrochim. Acta* **46** 3457–66

[236] Kavan L, Yum J-H and Graetzel M 2014 Graphene-based cathodes for liquid-junction dye sensitized solar cells: Electrocatalytic and mass transport effects *Electrochim. Acta* **128** 349–59

[237] Fabregat-Santiago F, Bisquert J, Palomares E, Otero L, Kuang D, Zakeeruddin S M and Grätzel M 2007 Correlation between Photovoltaic Performance and Impedance Spectroscopy of Dye-Sensitized Solar Cells Based on Ionic Liquids *J. Phys. Chem. C* **111** 6550–60

[238] Kavan L, Yum J-H, Nazeeruddin M K and Grätzel M 2011 Graphene Nanoplatelet Cathode for Co(III)/(II) Mediated Dye-Sensitized Solar Cells *ACS Nano* **5** 9171–8

[239] Mezzetti A, Balandeh M, Luo J, Bellani S, Tacca A, Divitini G, Cheng C, Ducati C, Meda L, Fan H. and Di

Fonzo F 2018 Hyperbranched TiO2 –CdS nano-heterostructures for highly efficient photoelectrochemical photoanodes *Nanotechnology* **29** 335404

[240]  Passoni L, Fumagalli F, Perego A, Bellani S, Mazzolini P and Di Fonzo F 2017 Multi-layered hierarchical nanostructures for transparent monolithic dye-sensitized solar cell architectures *Nanotechnology* **28** 245603

[241]  Gong F, Wang H and Wang Z-S 2011 Self-assembled monolayer of graphene/Pt as counter electrode for efficient dye-sensitized solar cell *Phys. Chem. Chem. Phys.* **13** 17676–82

[242]  Kavan L, Yum J H and Grätzel M 2011 Optically Transparent Cathode for Dye-Sensitized Solar Cells Based on Graphene Nanoplatelets *ACS Nano* **5** 165–72

[243]  Yum J-H, Baranoff E, Kessler F, Moehl T, Ahmad S, Bessho T, Marchioro A, Ghadiri E, Moser J-E, Yi C, Nazeeruddin M K and Grätzel M 2012 A cobalt complex redox shuttle for dye-sensitized solar cells with high open-circuit potentials *Nat. Commun.* **3** 631

[244]  Idígoras J, Guillén E, Ramos F J, Anta J A, Nazeeruddin M K and Ahmad S 2014 Highly efficient flexible cathodes for dye sensitized solar cells to complement Pt@TCO coatings *J. Mater. Chem. A* **2** 3175-3181

[245]  Lee C P, Li C T and Ho K C 2017 Use of organic materials in dye-sensitized solar cells *Mater. Today* **20** 267-283